\documentclass[a4paper]{article}

\usepackage{amssymb,amsfonts,amsthm}%,psfig}
\usepackage{graphicx}
\usepackage{subfigure}
\usepackage{amssymb}
\usepackage{amsthm}
\usepackage{amsmath}
\usepackage{bm}             %% boldmath-command: \bm{}
\usepackage[mathscr]{eucal}

\usepackage{cancel}

\usepackage{psfrag}

\usepackage{multirow}

\usepackage{wasysym}

\usepackage{color}

\def\pmb#1{\setbox0=\hbox{$#1$}%
  \kern-.025em\copy0\kern-\wd0
  \kern.05em\copy0\kern-\wd0
  \kern-.025em\raise.0433em\box0}
\def\pmbs#1{\setbox0=\hbox{$\scriptstyle #1$}%
  \kern-.0175em\copy0\kern-\wd0
  \kern.035em\copy0\kern-\wd0
  \kern-.0175em\raise.0303em\box0}
\def\be{\begin{equation}}
\def\ee{\end{equation}}
\def\bea{\begin{eqnarray}}
\def\eea{\end{eqnarray}}
\def\lb{\label}

\newcommand{\nt}{\tilde{N}}
\newcommand{\Htilde}{\tilde{\mathscr{H}}}
\newcommand{\Omegak}{\Omega_{\mathrm{k}}}

\newcommand{\cp}{c_{\mathrm{p}}}
\newcommand{\cs}{c_{\mathrm{s}}}

\def\cg{{\cal G}}

\def\bom{\mbox{\boldmath $\omega$}}

\def\bna{\mbox{\boldmath $\nabla$}}

\def\ptl{\partial}

\def\hsp5{\hspace{5mm}}
\def\case#1/#2{\textstyle\frac{#1}{#2}}

\theoremstyle{plain}

\theoremstyle{remark}

\newtheorem*{remark}{Remark}
\newcommand{\weg}{\:\,}
\DeclareMathOperator{\diag}{diag}
\DeclareMathOperator{\tr}{tr}

\newcommand{\Ug}{U_{\hspace{-0.1em}\mathrm{g}}}
\newcommand{\rg}{r_{\hspace{-0.1em}\mathrm{g}}}
\newcommand{\Uf}{U_{\mathrm{f}}}
\newcommand{\rf}{r_{\mathrm{f}}}
\renewcommand{\vector}[1]{\bm{#1}}
\newcommand{\defeq}{\mathrel{\mathop:}=}

\newcommand{\textfrac}[2]{{\textstyle \frac{#1}{#2}}}

\title{\sc Monotonic functions in Bianchi models: Why they exist and how to find them}
\author{ \\
{\Large\sc J.\ Mark Heinzle}\thanks{Electronic address:
{\tt mark.heinzle@univie.ac.at}} \\[1ex]
Gravitational Physics, Faculty of Physics, \\
University of Vienna, A-1090 Vienna, Austria
\and \\
{\Large\sc Claes Uggla}\thanks{Electronic address:
{\tt claes.uggla@kau.se}} \\[1ex]
Department of Physics, \\
University of Karlstad, S-651 88 Karlstad, Sweden \\[2ex] }

\date{\normalsize{December 01, 2009}}

\begin{document}
%%%%%%%%%%%%%%%%%%%%%%%%%%%%%%%%%%%%%%%%%%%%%%%%%%%%%%%%%%%%%%%%%%%
\maketitle
%\sloppy
%\doublespace

%%%%%%%%%%%%%%%%%%%%%%%%%%%%%%%%%%%%%%%%%%%%%%%%%%%%%%%%%%%%%%%%%%%
\begin{abstract}
%%%%%%%%%%%%%%%%%%%%%%%%%%%%%%%%%%%%%%%%%%%%%%%%%%%%%%%%%%%%%%%%%%%

All rigorous and detailed dynamical results in Bianchi cosmology
rest upon the existence of a hierarchical structure of
conserved quantities and monotonic functions. In this paper we
uncover the underlying general mechanism and
derive this hierarchical structure from the
scale-automorphism group for an illustrative example, vacuum
and diagonal class~A perfect fluid models. First,
kinematically, the scale-automorphism group leads to a reduced
dynamical system that consists of a hierarchy of
scale-automorphism invariant sets. Second, we show that,
dynamically, the scale-automorphism group results in
scale-automorphism invariant monotone functions and conserved
quantities that restrict the flow of the reduced dynamical
system.

%%%%%%%%%%%%%%%%%%%%%%%%%%%%%%%%%%%%%%%%%%%%%%%%%%%%%%%%%%%%%%%%%%%
\end{abstract}
%%%%%%%%%%%%%%%%%%%%%%%%%%%%%%%%%%%%%%%%%%%%%%%%%%%%%%%%%%%%%%%%%%%

\centerline{\bigskip\noindent PACS number(s): 04.20.-q, 98.80.Hw,
98.80.Dr, 04.20.Jb} \vfill
\newpage

%%%%%%%%%%%%%%%%%%%%%%%%%%%%%%%%%%%%%%%%%%%%%%%%%%%%%%%%%%%%%%%%%%%
\section{Introduction}
%%%%%%%%%%%%%%%%%%%%%%%%%%%%%%%%%%%%%%%%%%%%%%%%%%%%%%%%%%%%%%%%%%%

Spatially homogeneous Bianchi cosmology has been a popular
subject in general relativity ever since it was introduced by
Taub in 1951~\cite{taub51}. At first, because it made possible
the study of effects of nonlinear anisotropic perturbations of
spatially homogeneous and isotropic FRW models. More recently,
a new context for the dynamics of Bianchi cosmology emerged,
because it was realized that one can conformally rescale the
Einstein field equations so that the spatially homogeneous
equations occur on the boundary of the full state space of
general relativity---the so-called silent boundary---with
spatial coordinates appearing as an index set, which yields a
building block for %our understanding of
the detailed structure of certain special as well as generic
spacelike
singularities~\cite{uggetal03,rohugg05,andetal05,gar04,limetal06,heietal09}.

In Bianchi cosmology the space-time manifold $M$ is regarded as
a parameterized set of copies of a three-dimensional real Lie group
$G$ %$M=\{G(x^0)|x^0\in \mathbb{R}\}$,
%where $G$
that acts as a transformation group on $M$ with
three-dimensional spacelike orbits, which form a geodesically
parallel family of spatially homogeneous time slices, see,
e.g.,~\cite{waiell97,jan01} and references therein. The metric
on $M$ is defined by
\be ^4{\bf g} = -N^2(x^0) \,dx^0 \otimes dx^0 +
g_{\alpha\beta}(x^0)\:(\hat{\bom}^\alpha +
N^\alpha\,dx^0)\otimes (\hat{\bom}^\beta + N^\beta\,dx^0)
\qquad (\alpha,\beta=1,2,3), \ee
where $\{\hat{\bom}^\alpha\}$ is a left-invariant co-frame
on $G$ dual to a left-invariant spatial frame
$\{\hat{\vector{e}}_\alpha\}$. This frame is a basis of the Lie algebra
with structure constants $\hat{C}^\alpha{}_{\beta\gamma}$, i.e.,
\begin{subequations}\label{stucconsts}
\begin{equation}
[\hat{\vector{e}}_\beta,\hat{\vector{e}}_\gamma] =
\hat{C}^\alpha{}_{\beta\gamma}\,\hat{\vector{e}}_\alpha \qquad
\text{or, equivalently,} \qquad d\hat{\bom}^\alpha =
-\textfrac{1}{2}\hat{C}^\alpha{}_{\beta\gamma}\,\hat{\bom}^\beta\wedge
\hat{\bom}^\gamma .
\ee
The structure constants can be decomposed as
follows~\cite{estetal68}:
\be\lb{Cdecomp} \hat{C}^\alpha{}_{\beta\gamma} =
\epsilon_{\beta\gamma\delta}\,\hat{n}^{\alpha\delta} +
\hat{a}_\sigma\,\delta^{\sigma\alpha}_{\beta\gamma} ,\qquad
\hat{a}_\sigma=\textfrac{1}{2} \hat{C}^\alpha{}_{\sigma\alpha} . \ee
\end{subequations}
The Bianchi models are divided into two main classes: The class
A models for which $\hat{a}_\alpha=0$, and the class B models
for which $\hat{a}_\alpha\neq 0$. All class A models admit a
Hamiltonian description while this is only the case for a few
class B models, see, e.g.,~\cite{jan01}. Although our results
do not depend on Hamiltonian methods, a Hamiltonian approach
simplify things. In this paper we will therefore be concerned
with class A models.

The foundation for basically all rigorous results on the
dynamics of Bianchi cosmologies is the existence of an
increasingly restrictive hierarchy of monotonic functions and
conserved quantities, which is associated with a hierarchy of
Lie and source contractions. But why do useful monotonic
functions and conserved quantities exist at all, and how does
one find them? The purpose of this paper is to reveal and
exploit the underlying general mechanism, namely the
scale-automorphism group, by thoroughly examining a specific
example: We will restrict ourselves to the class~A diagonal
vacuum and orthogonal perfect fluid models, for which the fluid
4-velocity is orthogonal to the spatially homogeneous symmetry
surfaces. In addition, we will assume that the perfect fluid
satisfies a barotropic equation of state, $p=p(\rho)$, and we will
focus on linear equations of state $p=w\rho$ with $-1<w<1$,
$w\neq -1/3$, where $p$ and $\rho$ is the pressure and energy
density, respectively. In this paper, we will
\textit{derive}, \textit{from first principles} (i.e., from the
scale-automorphism group), the structure that is necessary to
describe the dynamics of the models under consideration,
%(the
%hierarchy of conserved quantities and monotone functions),
a structure that is the basis of every available theorem in
this context~\cite{waiell97,rin01,heiugg09a}:
A hierarchy of conserved quantities and monotone functions.

Some of the ideas in this paper have precursors in work by
Uggla in~\cite[Chapter 10]{waiell97}, which was used in the
proofs of some of the theorems in~\cite{waiell97} and in the
proofs of the Mixmaster attractor theorem
in~\cite{rin00,rin01,heiugg09a,heiugg09b}. The analysis
of~\cite[Chapter 10]{waiell97} in turn rests upon earlier work
in~\cite{jan79,rosetal90a,rosetal90b,uggetal91,jan01} and
references therein. Furthermore, some of our results were
inspired by material presented in a talk by Uggla at the Newton
institute in 2005. However, here we develop, for the first
time, the complete picture in full detail. Moreover, we use
different and more efficient general techniques than in the
precursor material; this in turn sets the stage for
developments as regards more general and complicated
situations.

The outline of the paper is as follows. In the next section we
give the Hamiltonian equations of the present class A models
and derive the so-called reduced Hubble-normalized dynamical
system which has been the framework for our detailed
understanding of class A vacuum and orthogonal perfect fluid
cosmology~\cite{waiell97,rin00,rin01,heiugg09b}; however, see,
e.g.,~\cite{heiugg09a,heietal05} for other useful variables. In
Section~\ref{Sec:scaleaut} we present the scale-automorphism
group. In Section~\ref{Sec:scaleautinv} we show that the
reduced Hubble-normalized dynamical system is a kinematical
consequence of the scale-automorphism group; it is a
self-contained system for the scale-automorphism invariant,
i.e., scale and gauge invariant, degrees of freedom.
Section~\ref{Sec:Hamstruc} contains Hamiltonian considerations
of a general nature. For a class of Hamiltonians that
encompasses the class~A Hamiltonians we derive conserved
quantities and monotone functions, and analyze the conditions
under which such functions are invariant under a Lie group of
transformations (such as the scale-automorphism group). These
results are subsequently applied in Section~\ref{Sec:dyncon}:
Using the scale-automorphism group we derive conserved
quantities and monotone functions in a step-by-step manner for
each Bianchi model; these objects are expressed in terms of
the state vector of the reduced Hubble-normalized dynamical
system of Section~\ref{Hamdyn}. We conclude with a discussion
in Section~\ref{discussion} where we argue that the present
work is just an illustration of a phenomenon with much broader
ramifications, e.g., we discuss the effects of the
scale-automorphism group in the context of the Einstein-Vlasov
system.

%%%%%%%%%%%%%%%%%%%%%%%%%%%%%%%%%%%%%%%%%%%%%%%%%%%%%%%%%%%%%%%%%%%
\section{Hamiltonian approach and dynamical systems framework}\label{Hamdyn}
%%%%%%%%%%%%%%%%%%%%%%%%%%%%%%%%%%%%%%%%%%%%%%%%%%%%%%%%%%%%%%%%%%%

In this section we use the Hamiltonian description of class~A
Bianchi cosmology to derive the Hubble-normalized dynamical
systems formulation. We consider the vacuum and orthogonal
perfect fluid case, for which one can choose an adapted frame that
simultaneously diagonalizes the metric and the matrix
$\hat{n}^{\alpha\beta}$ of~\eqref{stucconsts}, i.e.,
\begin{subequations}\label{metric}
\begin{gather}
\label{diagmetric}
{}^4\mathbf{g}
= -N^2\, d
x^0 \otimes d x^0 + g_{11}\, \hat{\bom}^1\otimes \hat{\bom}^1 +
g_{22}\, \hat{\bom}^2\otimes \hat{\bom}^2 + g_{33}\,
\hat{\bom}^3\otimes \hat{\bom}^3 \:,
\\[0.5ex]
d\hat{\bom}^1
=  -\hat{n}_1 \, \hat{\bom}^2\wedge
\hat{\bom}^3 ,\quad d\hat{\bom}^2  =  -\hat{n}_2 \,
\hat{\bom}^3\wedge \hat{\bom}^1 ,\quad d\hat{\bom}^3  =
-\hat{n}_3 \, \hat{\bom}^1\wedge \hat{\bom}^2 \:.
\end{gather}
\end{subequations}
see, e.g.,~\cite{waiell97}.
The structure constants $\hat{n}_1$, $\hat{n}_2$, $\hat{n}_3$
represent the symmetry group; a classification of the
various class~A models given in Table~\ref{classAmodels}.
\begin{table}
\begin{center}
\begin{tabular}{|c|ccc|}
\hline Bianchi type & $\hat{n}_\alpha$ &  $\hat{n}_\beta$ & $\hat{n}_\gamma$ \\ \hline
I & $0$ & $0$ & $0$ \\
II & $+$ & $0$ & $0$ \\
$\mathrm{VI}_0$ & $+$ & $-$ & $0$ \\
$\mathrm{VII}_0$ & $+$& $+$ & $0$ \\
$\mathrm{VIII}$ & $+$& $+$& $-$ \\
$\mathrm{IX}$ & $+$& $+$& $+$ \\\hline
\end{tabular}
\caption{The Bianchi types that belong to class A are
characterized by different relative signs of the structure constants
$(\hat{n}_\alpha, \hat{n}_\beta, \hat{n}_\gamma)$, where
$(\alpha\beta\gamma)$ is any permutation of $(123)$. In
addition to the above representations there exist equivalent
representations associated with an overall change of sign of
the structure constants; e.g., another type IX representation
is $(---)$.} \label{classAmodels}
\end{center}
\end{table}
It is convenient to represent the metric~\eqref{diagmetric} as
\begin{equation}\label{diagmetric2}
{}^4\mathbf{g} = -g\,\tilde{N}^2\, d x^0 \otimes d x^0 +
e^{2\beta^1}\hat{\bom}^1\otimes \hat{\bom}^1 +
e^{2\beta^2}\hat{\bom}^2\otimes \hat{\bom}^2 +
e^{2\beta^3}\hat{\bom}^3\otimes \hat{\bom}^3  \:,
\tag{\ref{diagmetric}${}^\prime$}
\end{equation}
where $g$ is the determinant of the spatial metric, i.e.,
$g = \det g = \exp [2(\beta^1 + \beta^2 + \beta^3)]$.

%-----------------------------------------------------------------
\subsection{Hamiltonian equations}\label{Ham}
%-----------------------------------------------------------------

The scalar Hamiltonian (density) is given
by~\cite{waiell97,jan01}
\begin{equation}\label{Hami1st}
\Htilde = 2\,N \sqrt{g} \:(n^a n^b G_{ab} - n^a n^b T_{ab}) =
-\tilde{N}g \,\Big( (\tr k)^2 - k^\alpha_{\weg \beta} k^\beta_{\weg \alpha}
 + {}^3\!R - 2 \rho \Big) = 0\:,
\end{equation}
where $n^a$ is the unit normal vector field of the spatially
homogeneous foliation; $G_{ab}$ is the Einstein tensor, and
$T_{ab}$ is the stress-energy tensor --- we use units such that
Newton's gravitational constant $G$ and the speed of light $c$
are given by $8 \pi G =1$ and $c =1$. The expression
$k_{\alpha\beta}$ denotes the second fundamental form of the
spatial hypersurfaces; ${}^3\!R$ is the scalar curvature of the
three-metric $g_{\alpha\beta}$, and $\rho$ is the energy density. % (w.r.t.\ $n^a$).
The relation $\Htilde = 0$ is the Hamiltonian constraint.

In the special case~\eqref{diagmetric}
%of diagonal class~A vacuum and orthogonal
%perfect fluid models
we have %the relation
$\partial_0 g_{\alpha\beta} = -2 N k_{\alpha\beta}$, where $\partial_0 =
\partial/\partial x^0$, and thus
\begin{equation}\label{kinbetadot}
k^\alpha_{\weg \beta} = g^{\alpha\gamma}k_{\gamma\beta}=
\diag\big(k^1_{\weg 1}, k^2_{\weg 2}, k^3_{\weg 3}\big)
= -N^{-1} \diag\big( \partial_0 \beta^1, \partial_0 \beta^2, \partial_0 \beta^3\big)\:.
\end{equation}
Expressing~\eqref{Hami1st} in terms of $\beta^\delta$ and
$\partial_0 \beta^\delta$, $\delta = 1,2,3$,
cf.~\eqref{kinbetadot}, allows us to apply the standard formalism
to obtain the momenta $\pi_\delta$ that are canonically
conjugate to $\beta^\delta$. Let $(\alpha\beta\gamma)$ be a
(fixed) permutation of $(123)$, %(i.e., there is no summation
%over repeated indices);
then
\begin{equation}\label{piink}
\pi_\alpha = -2 N^{-1} \sqrt{g} \,\big( \partial_0 \beta^\beta + \partial_0\beta^\gamma \big) =
2 \sqrt{g} \,\big(k^\beta_{\weg \beta} + k^\gamma_{\weg \gamma} \big)\:,
\qquad
k^\alpha_{\weg\alpha} = -\textfrac{1}{4}\, g^{-1/2}\,
(\pi_\alpha - \pi_\beta - \pi_\gamma)\:,
\end{equation}
%
%with the inversion
%%
%\begin{equation}\label{kpi}
%k^\alpha_{\weg\alpha} = -\textfrac{1}{4}\, g^{-1/2}\,
%(\pi_\alpha - \pi_\beta - \pi_\gamma)\:,
%\end{equation}
%%
cf.~the Hamiltonian equation~\eqref{pidef} below.

A key step is to
introduce the
so-called minisuperspace metric $\cg_{\alpha\beta}$
and its inverse $\cg^{\alpha\beta}$ by
\begin{equation}\label{minisupmetric}
\cg_{{\alpha} \beta} =
\begin{pmatrix}
0 & -1 & -1 \\
-1 & 0 & -1 \\
-1 & -1 & 0
\end{pmatrix}
;\qquad\quad
\cg^{{\alpha} \beta} = \textfrac{1}{2}\begin{pmatrix}
1 & -1 & -1 \\
-1 & 1 & -1 \\
-1 & -1 & 1
\end{pmatrix}\:.
\end{equation}
The signature of $\cg_{\alpha\beta}$ is $(-++)$, and hence
$\cg_{\alpha\beta}$ is a $(2+1)$-dimensional minisuperspace
Minkowski metric.
Based on~\eqref{piink} and~\eqref{minisupmetric} we
have
\begin{equation}\label{threecurv}
k^\alpha_{\weg \beta} k^\beta_{\weg \alpha}- (\tr k)^2 =
\textfrac{1}{4}\, g^{-1} \,\cg^{{\gamma} \delta} \pi_{\gamma} \pi_{\delta}
\quad\text{ and }\quad
{}^3\!R  = -g^{-1} \mathcal{G}^{\gamma\delta} (\hat{n}_\gamma e^{2 \beta^\gamma})
(\hat{n}_\delta e^{2 \beta^\delta}) \:,
\end{equation}
where we sum over $\gamma$, $\delta$.
Like ${}^3\!R$, the energy density $\rho$ in~\eqref{Hami1st}
can be expressed in terms of the metric variables
$\beta^\delta$, $\delta=1,2,3$, which is because, in principle,
the conservation law $\bna_a T^{ab} =0$ can be solved for
barotropic equations of state $p=p(\rho)$; in the present case
we obtain
\begin{equation}\label{rhoeq}
\frac{d\rho}{\rho + p(\rho)} =  \frac{d\rho}{\rho(1 +
w(\rho))} = -\frac{dg}{2g}\:;
\end{equation}
therefore, $\rho=\rho(g)$; the function $\rho(g)$ is monotonically decreasing
if the weak energy condition is strictly satisfied,
i.e., if $\rho>0$ and $\rho + p> 0$ (i.e., $w > -1$). For a
linear equation of state $p=w\rho$ with $w=\mathrm{const}$,~\eqref{rhoeq}
yields
\begin{equation}\label{rhoeq2}
\rho = \rho_0 \,g^{-(1+w)/2} = \rho_0\,e^{-(1+w)(\beta^1 + \beta^2 + \beta^3)} \:,
\tag{\ref{rhoeq}${}^\prime$}
\end{equation}
where $\rho_0$ is a constant of integration.

Making use of the above results, the
Hamiltonian~\eqref{Hami1st} reads
\begin{equation}\label{Hamiltonian}
\Htilde = \nt\,\mathscr{H} =
\tilde{N} \left( \textfrac{1}{4}
\cg^{{\gamma} \delta} \pi_{\gamma} \pi_{\delta} - {}^{3}\!R\,g  + 2\rho\,g\right)
= \tilde{N} \cg^{\gamma \delta}
\left[\textfrac{1}{4}\pi_{\gamma}\pi_{\delta} +
(\hat{n}_\gamma e^{2 \beta^\gamma})(\hat{n}_\delta e^{2 \beta^\delta}) +
\textfrac{4}{3} \rho\, g\,\delta_{\gamma\delta}\right] \:.
\end{equation}
We split $\mathscr{H}$ into a kinetic part $T$,
a gravitational potential $\Ug$, and a fluid potential $\Uf$, i.e.,
\begin{subequations}\label{hamspecific}
\begin{align}
\label{hamsplit}
\Htilde & = \nt\,\mathscr{H} =
\nt(T + \Ug + \Uf) =0\:, \quad\text{where} \\[0.5ex]
\label{kineticpart}
&T = \textfrac{1}{4}\cg^{{\gamma} \delta}
\pi_{\gamma}\pi_{\delta}\:, \\[0.5ex]
\label{gravipot}
&\Ug = \cg^{{\gamma} \delta}
(\hat{n}_\gamma e^{2 \beta^\gamma})(\hat{n}_\delta e^{2 \beta^\delta})\:,\\[0.5ex]
\label{fluidpot}
& \Uf = 2\rho g = 2\rho_0 e^{(1-w)(\beta^1 + \beta^2 + \beta^3)}\:.
\end{align}
\end{subequations}
Note that the second expression in~\eqref{fluidpot}
requires a
linear equation of state, i.e., $w=\mathrm{const}$.
%where $\Uf$ depends only on the sum $\beta^1 + \beta^2 +
%\beta^3$

If we regard $\nt$ as an independent variable then variation
w.r.t.\ $\nt$ yields the Hamiltonian constraint
$\mathscr{H}=0$, and we obtain the %following
Hamiltonian equations
\begin{subequations}\label{hameq}
\begin{align}
\label{pidef}
\frac{d\beta^\alpha}{d x^0} & =
\frac{\partial \Htilde}{\partial \pi_\alpha} =
%\nt\frac{\partial {\cal H}}{\partial \pi_\alpha} =
\textfrac{1}{4}\, \nt\, (\pi_\alpha - \pi_\beta - \pi_\gamma) \:, \\[0.5ex]
\label{hamm2}
\frac{d\pi_\alpha}{d x^0} &=
-\frac{\partial \Htilde}{\partial \beta^\alpha} =
%-\nt\frac{\partial {\cal H}}{\partial \beta^\alpha} =
-2\nt \left[\hat{n}_\alpha e^{2 \beta^\alpha}
\big(\hat{n}_\alpha e^{2\beta^\alpha} - \hat{n}_\beta e^{2\beta^\beta}
- \hat{n}_\gamma e^{2\beta^\gamma}\big)
+  (1-w)\rho_0e^{(1-w)(\beta^1 + \beta^2 + \beta^3)}\right]  \:.
\end{align}
\end{subequations}
In~\eqref{hameq}, $(\alpha\beta\gamma)$ is a cyclic
permutation of $(123)$ and no sums are taken over repeated
indices.
%Alternatively, we can regard $\nt$ as a function of
%$\beta^\alpha$ and $\pi_\alpha$ and impose the Hamiltonian
%constraint by hand; this yields
%%
%%\begin{subequations}\label{hameq2}
%\begin{align*}
%\frac{d\beta^\alpha}{d x^0} &= \frac{\partial \Htilde}{\partial
%\pi_\alpha} = \mathscr{H} \frac{\partial \nt}{\partial
%\pi_\alpha} + \nt\frac{\partial \mathscr{H}}{\partial
%\pi_\alpha} = \nt\frac{\partial \mathscr{H}}{\partial
%\pi_\alpha} = \text{\eqref{pidef}}
%\:, \\
%\frac{d\pi_\alpha}{d x^0} &= -\frac{\partial \Htilde}{\partial
%\beta^\alpha} =  -\mathscr{H}\frac{\partial \nt}{\partial
%\beta^\alpha} - \nt\frac{\partial \mathscr{H}}{\partial
%\beta^\alpha} = - \nt\frac{\partial \mathscr{H}}{\partial
%\beta^\alpha}
%= \text{\eqref{hamm2}}
%\:,
%\end{align*}
%%\end{subequations}
%
%because $\mathscr{H}=0$; i.e., on shell the two approaches
%lead to the same result.
Note that the Hamiltonian momentum constraints are identically zero and
thus
automatically satisfied for the present models; this is because
both the Einstein tensor of a diagonal class~A metric and the
the stress-energy tensor for an orthogonal perfect fluid are diagonal.

It is useful to introduce additional variables, $\beta^0$ and
$\pi_0$, which are part of the Misner parameterization of the
metric variables~\cite{mis69a,mis69b},
%
%\begin{subequations}\label{def0}
%\begin{gather}
%\label{pinpi} \beta^0 =  \textfrac{1}{3}(\beta^1 + \beta^2 +
%\beta^3)\:, \qquad\quad \pi_0 = \pi_1 + \pi_2 + \pi_3 \\[0.5ex]
%%\label{g0} g = e^{6\beta^0}\:,  \quad  \rho = \rho(\beta^0)
%%= \rho_0 \,e^{-3(1+w) \beta^0}\:,\quad
%\quad \Rightarrow \quad
%\Uf = \Uf(\beta^0) = 2
%\rho g = 2 \rho_0 \,e^{-3(1-w) \beta^0}\:.
%\end{gather}
%\end{subequations}
%
%\begin{subequations}\label{def0}
\begin{equation}
\label{pinpi}
\beta^0 =  \textfrac{1}{3}(\beta^1 + \beta^2 +
\beta^3)\:, \qquad\quad \pi_0 = \pi_1 + \pi_2 + \pi_3\:,
\end{equation}
and to express~\eqref{fluidpot} in terms of $\beta^0$, i.e.,
$\Uf = \Uf(\beta^0) = 2 \rho g = 2 \rho_0 \,e^{-3(1-w) \beta^0}$.

\subsection{The Hubble-normalized dynamical systems approach}\label{Hub}
%-----------------------------------------------------------------

The main idea of the Hubble-normalized dynamical systems
approach to Bianchi cosmology is to `factor out' the expansion
(or, equivalently, the Hubble variable) and decouple the gauge
degrees of freedom from the `essential' dynamics. The Hubble
variable $H$ (which is not to be confused with the Hamiltonian
$\mathscr{H}$) is proportional to the expansion $\theta$ of the
congruence of geodesics orthogonal to the symmetry surfaces and
thus to the mean curvature $\tr k$, i.e.,
$H = \textfrac{1}{3}\,\theta = \textfrac{1}{3} \,N\, \frac{d}{d x^0} \log \sqrt{g} =
 -\textfrac{1}{3} \,\tr k$. Using~\eqref{piink} and $\pi_0 =
\pi_1 + \pi_2 + \pi_3$, cf.~\eqref{pinpi}, we see that
\begin{equation}\label{Hinp0}
H = -\textfrac{1}{3} \,\tr k =
%\textfrac{1}{3} \,N\, \frac{d}{d x^0} \log \sqrt{g} =
%\textfrac{1}{3}\,\tilde{N}\, \frac{d}{d x^0} \sqrt{g} = -\textfrac{1}{3} \,\tr k =
- \textfrac{1}{12} \,g^{-1/2} \:\pi_0 = -\textfrac{1}{12}\, e^{-3\beta^0}\,\pi_0\:,
\end{equation}
%
%recall that $g = e^{2 (\beta^1+\beta^2+\beta^3)} = e^{6 \beta^0}$.
One is primarily interested in expanding cosmological models,
i.e., $H > 0$ ($\leftrightarrow$ $\pi_0 < 0$).
%$\theta>0$ (which corresponds to $H
%> 0$ and $\pi_0 < 0$).
The shear
$\sigma_{\alpha\beta}=\diag(\sigma_1,\sigma_2,\sigma_3)$
of the congruence of
geodesics orthogonal to the symmetry surfaces is
\begin{equation}\label{sigpi}
\sigma_\alpha = {-}k^\alpha_{\weg \alpha} + \textfrac{1}{3}\, \tr k =
{-}k^\alpha_{\weg \alpha} - H =
\textfrac{1}{2}\, g^{-1/2}\,\left(\pi_\alpha -\textfrac{1}{3}\pi_0\right) =
\textfrac{1}{2}\, e^{-3\beta^0}\,\left(\pi_\alpha -\textfrac{1}{3}\pi_0\right)\:,
\end{equation}
see~\eqref{piink}; $\sigma_1 + \sigma_2 + \sigma_3 = 0$.
The Hubble-normalized dynamical systems approach is based
on the definition of dimensionless variables.
%The
%definition of the dimensionless Hubble-normalized dynamical
%systems variables is based on the metric variables
%$\big(g_{\alpha\alpha}, k^\alpha_{\weg \alpha}\big)$, $\alpha
%=1,2,3$, where $(k^\alpha_{\weg \alpha})$ is split into $(H,
%\sigma_\alpha)$, or, equivalently, on the Hamiltonian variables
%$(\beta^\alpha, \pi_\alpha)$.
Let $(\alpha\beta\gamma)$ be a
permutation of $(123)$; then
\begin{subequations}\label{sigNdef}
\begin{align}
\label{sigdef}
\Sigma_\alpha &= \frac{\sigma_\alpha}{H} =
({-}6) \left(\frac{\pi_\alpha}{\pi_0} - \frac{1}{3} \right) =
2 \,\pi_0^{-1}\big[(\pi_\beta - \pi_\alpha) + (\pi_\gamma - \pi_\alpha)\big]\:, \\[1ex]
\label{Ndef}
N_\alpha & = \hat{n}_\alpha \frac{g_{\alpha\alpha}}{\sqrt{g} \,H} =
%\hat{n}_\alpha \frac{e^{2\beta^\alpha}}{\sqrt{g}\,H} =
-12\, \hat{n}_\alpha \, \pi_0^{-1} e^{2\beta^\alpha}\:,
\end{align}
\end{subequations}
cf.~\cite{waiell97}, where we have used~\eqref{Hinp0}
and~\eqref{sigpi}; clearly, $\Sigma_1 + \Sigma_2 + \Sigma_3 =
0$; we note that $N_\alpha = 0$ when $\hat{n}_\alpha = 0$.
%While $H$ is the normalization factor for the metric variables,
%this role is taken over by ${-\pi_0}$ in the Hamiltonian
%approach.
Eq.~\eqref{sigdef} implies
\begin{equation}\label{sigdefstrich}
\pi_\delta=\textfrac{1}{6}(2-\Sigma_\delta)\,\pi_0 \qquad\quad(\delta = 1,2,3),
\tag{\ref{sigdef}${}^\prime$}
\end{equation}
which is consistent with $\pi_1 + \pi_2 + \pi_3 = \pi_0$, since
$\Sigma_1 + \Sigma_2 + \Sigma_3 = 0$.

Since the variables $N_\alpha$, $\alpha = 1,2,3$, incorporate
the structure constants $\hat{n}_\alpha$, the variable
transformation between the original variables and
$(H,\Sigma_\alpha, N_\alpha)$ is one-to-one only for Bianchi
types~VIII and~IX (where $\hat{n}_\alpha \neq 0$ $\forall
\alpha$). For the lower Bianchi types (I, II, $\mathrm{VI}_0$,
$\mathrm{VII}_0$) we may define
\begin{equation}\lb{Mdelta}
M_\delta  = \frac{g_{\delta\delta}}{\sqrt{g} \,H}  =
-12 \, \pi_0^{-1} e^{2\beta^\delta} \qquad\quad(\delta = 1,2,3)\,.
%\tag{\ref{Ndef}${}^\prime$}
\end{equation}
To reconstruct the original variables (metric or Hamiltonian)
from $(\Sigma_\alpha, N_\beta)$ we have to add $H$ and one
variable $M_\delta$ for each missing variable $N_\delta$ (when
$\hat{n}_\delta =0$).
%However, in Section~\ref{Sec:scaleautinv}
%we show that $M_\delta$ ($\hat{n}_\delta = 0$) are associated
%with the gauge degrees of freedom while $H$ carries the scale
%degree of freedom, and thus the set of variables
%$(\Sigma_\alpha, N_\beta)$, where $\beta$ is such that
%$\hat{n}_\beta\neq 0$, carries the essential dynamical content
%in all cases.
%We will refer to the system that
%describes the evolution of these variables as the
%\textit{reduced\/} Hubble-normalized system,
%cf.~\eqref{Hdynsys}.

In addition to the variables~\eqref{sigNdef} we define the
Hubble-scaled energy density $\Omega = \rho/(3 H^2)$ and the
Hubble-scaled spatial curvature scalar ${-2\Omegak}= {}^3\!R/(3
H^2)$; hence
\begin{subequations}
\begin{align}\label{Omegadef}
\Omega & =  \frac{\rho}{3 H^2} =
\frac{48 \rho\, g}{\pi_0^2} = \frac{48 \rho_0 \,e^{-3(1+w)\beta^0}}{\pi_0^2}
=\frac{24 \Uf}{\pi_0^2} \:, \\
\nonumber
\Omegak & = -\frac{{}^3\!R}{6 H^2} =  {-}\frac{24\, {}^3\!R\, g}{\pi_0^2} =
\frac{24 \Ug}{\pi_0^2} =\textfrac{1}{6} \mathcal{G}^{\gamma\delta} N_\gamma N_\delta =
\textfrac{1}{12} \left[ N_1^2 + N_2^2 + N_3^2 - 2 \big(N_1 N_2 + N_1 N_3 + N_2 N_3 \big)\right].
\end{align}
\end{subequations}
Note that $\Omegak$ simplifies when a structure constant $\hat{n}_\alpha$
is zero (since then $N_\alpha=0$).
In particular, $\Omegak = 0$ in Bianchi type~I;
$\Omegak = \textfrac{1}{12} N_\alpha^2$ in type~II;
$\Omegak = \textfrac{1}{12} (N_\alpha - N_\beta)^2$ in
type~$\mathrm{VI}_0$ and~$\mathrm{VII}_0$.

Finally, we Hubble-normalize the tracefree part of the spatial three-curvature
and obtain
\begin{equation}\label{3Salpha}
{}^3\!S_\alpha  =
\frac{{}^3\!R^\alpha_{\weg \alpha} - \textfrac{1}{3} {}^3\!R}{H^2} =
\frac{144\,g\, ({}^3\!R^\alpha_{\weg \alpha} - \textfrac{1}{3}{}^3\!R)}{\pi_0^2}
= \textfrac{1}{3} \Big[ N_\alpha ( 2 N_\alpha - N_\beta - N_\gamma) - (N_\beta - N_\gamma)^2 \Big]\:,
\end{equation}
where we use~\eqref{threecurv} and
\begin{equation*}
{}^3\!R^\alpha_{\weg \alpha} = \frac{1}{2 g} \Big[ \hat{n}_\alpha^2 e^{4 \beta^\alpha} -
\big( \hat{n}_\beta e^{2 \beta^\beta} - \hat{n}_\gamma e^{2 \beta^\gamma}\big)^2 \Big]\:,
\qquad\quad (\alpha\beta\gamma) \in\big\{(123),(231),(312)\}\:.
\end{equation*}
%
%Instead of~\eqref{3Salpha} it is sometimes advantageous to use
%%
%\begin{equation}
%{}^{3}\!S_\alpha = N_\alpha(N_\alpha - N_\beta - N_\gamma) - 4\Omegak
%\qquad\quad (\alpha\beta\gamma) \in\big\{(123),(231),(312)\}\:.
%\tag{\ref{3Salpha}${}^\prime$}
%\end{equation}
%%

Apart from the Hubble-scaled variables and matter/curvature
quantities, we also introduce a scaled lapse $H N$, which we
set to one, i.e.,
\begin{equation}\label{tauchoice}
N= H^{-1} \quad \Leftrightarrow \quad \nt = -12\, \pi_0^{-1}\:,
\end{equation}
in order to obtain a scale-invariant (see
Section~\ref{Sec:scaleautinv}) time variable $x^0$, which we
denote by $\tau$. This results in $d\tau = d\beta^0$, since the
Hamiltonian equations~\eqref{hameq} % associated with~\eqref{Hamiltonianin0pm}
yield
\begin{equation}\label{taudef}
\frac{d\beta^0}{d \tau} = \frac{\partial\Htilde}{\partial \pi_0}
= - \textfrac{1}{12}\tilde{N}\,\pi_0 = 1\:.
\end{equation}

With this choice of time variable, the Hamiltonian
constraint~\eqref{Hamiltonian} can be written as
\begin{equation}\label{Hresc}
2 \pi_0^{-1} \Htilde = -24 \pi_0^{-2} \cg^{\gamma \delta}
\left[\textfrac{1}{4}\pi_{\gamma}\pi_{\delta} +
(\hat{n}_\gamma e^{2 \beta^\gamma})(\hat{n}_\delta e^{2 \beta^\delta}) \right] -
48 \pi_0^{-2} \rho\, g\, = 1 - \Sigma^2 - \Omegak - \Omega = 0\:,
\end{equation}
where $\Sigma^2 \defeq \textfrac{1}{6}
\big(\Sigma_1^2+\Sigma_2^2+\Sigma_3^2\big)$.

The Hamiltonian equations~\eqref{hameq} and the Hamiltonian
constraint $\Htilde = 0$ lead to the following \textit{reduced
Hubble-normalized dynamical system\/} of evolution and
constraint equations for the `essential' Hubble-normalized
variables $(\Sigma_\alpha, N_\beta)$ ($\hat{n}_\beta \neq 0$):
\vspace{0.5ex}
\begin{subequations}\label{Hdynsys}
\begin{align}
\label{evol}
\text{Evolution eqs.}\:
& \left\{
\begin{array}{ll}
   \Sigma_\alpha^\prime = -(2-q)\Sigma_\alpha - {}^{3}\!S_\alpha & \quad (\alpha=1,2,3)\:, \\[1ex]
    N_\beta^\prime =  (q+2\Sigma_\beta)\,N_\beta &\quad (\hat{n}_\beta \neq 0, \text{no sum over $\beta$})\,,
\end{array} \right. \\[1ex]
\label{constraints}
\text{Constraints} \:
&\left\{\begin{array}{l}
    0 = \Sigma_1 + \Sigma_2 + \Sigma_3  \,,
    \\[0.75ex]
    0 = 1 - \Sigma^2 - \Omegak - \Omega   \,.
\end{array}\right.
\end{align}
\end{subequations}
Here and henceforth, a prime denotes the derivative w.r.t.\
$\tau$. The quantity $q$ denotes the deceleration parameter,
%which is originally defined through $1 + q = H^{-1} H^\prime$;
which is given by
\begin{equation}\label{decel}
q = 2\Sigma^2 + \textfrac{1}{2}(1+3w)\Omega\:.
\end{equation}
Note that $2 - q = 2 \Omegak +\textfrac{3}{2} (1 -w)\Omega$.
In the system~\eqref{Hdynsys} we can use the Hamiltonian
constraint (Gauss constraint)
to globally solve for $\Omega$ according to $\Omega = 1 -
\Sigma^2 - \Omegak$; consequently, the system~\eqref{Hdynsys}
only involves $\Sigma_\alpha$ ($\alpha =1,2,3$),
the (non-zero) variables $N_\beta$ ($\hat{n}_\beta \neq 0$),
and $w$.

In addition to~\eqref{Hdynsys}, the Hamiltonian equations~\eqref{hameq} imply
the evolution equations
\begin{equation}\label{p0eq}
H^\prime = -(1+q)H \qquad\Leftrightarrow\qquad
\pi_0^\prime = -\frac{\partial\Htilde}{\partial\beta^0} = (2-q)\,\pi_0
\end{equation}
for the Hubble scalar $H$ and $\pi_0$.
Other equations of interest are the auxiliary equations for
$\Omega$, $M_\delta$, $\delta = 1,2,3$, and the equation for the Hamiltonian
variables $\beta^\alpha$ and $\pi_\alpha$, $\alpha =1,2,3$.
\begin{subequations}\label{auxeqs}
\begin{alignat}{2}
\label{Omegaeq}
& \Omega^\prime  = \big(2q - (1 + 3w)\big)\Omega\:, & \qquad &
M_\delta^\prime =  (q + 2\Sigma_\delta)\,M_\delta \qquad
(\hat{n}_\delta = 0, \text{no sum over $\delta$})\:,\\
\label{auxeq} & (\beta^\alpha)^\prime = 1 + \Sigma_\alpha\:, &
\qquad & \pi_\alpha^\prime  =
-\frac{\partial\Htilde}{\partial\beta^\alpha} =
\textfrac{1}{6}\,\pi_0 [N_\alpha(N_\alpha - N_\beta - N_\gamma)
+ 3(1-w)\Omega]\:.
\end{alignat}
\end{subequations}
%
%To obtain the latter equation one uses~\eqref{hameq} and the
%definitions of the Hubble-scaled variables and quantities.

Both in the vacuum case and for a perfect fluid with
a linear equation of state $w=\mathrm{const}$, the reduced
dynamical system~\eqref{Hdynsys} %is a self-contained system of
%autonomous equations
completely describes the dynamics of
Bianchi models of class~A;
the system~\eqref{Hdynsys}
contains \mbox{$[2 + \text{number of } \hat{n}_\beta \neq 0]$}
degrees of freedom in the perfect fluid case with $w = \mathrm{const}$,
which is in contrast to the Hamiltonian
problem which a priori involves six degrees of freedom
$\{\beta^\alpha, \pi_\alpha\}$.
Due to the decoupling of~\eqref{p0eq}
and~\eqref{auxeqs}, % decouple from~\eqref{Hdynsys},
one reconstructs the metric~\eqref{metric} from a solution of%
%the reduced system
~\eqref{Hdynsys} in a straightforward manner:
Integration of~\eqref{p0eq} yields $H$, which, together with
the solution of~\eqref{Hdynsys}, algebraically leads to the
metric via~\eqref{sigNdef} for Bianchi types~VIII and~IX; for
the lower Bianchi types one also has to integrate $M_\delta$
(when $\hat{n}_\delta=0$) by means of~\eqref{Omegaeq}, and then
use~\eqref{Mdelta} to obtain the metric component
$g_{\delta\delta}$. In Section~\ref{Sec:scaleautinv} we show
that the decoupling of $H$ and $M_\delta$ is due to the fact
that these variables are scale and gauge variables,
respectively.
 % we refer to
%Section~\ref{Sec:scaleautinv} for a comprehensive discussion.

In the following we derive from first principles monotone
functions and conserved quantities that restrict, or even determine,
the flow on
the state space of the reduced dynamical
system~\eqref{Hdynsys}. We begin by defining and discussing the
scale-automorphism group and its properties.

%i.e., from the scale-automorphism group.

%Let us finally remark that it is sometimes convenient to solve
%the constraint $\Sigma_1 + \Sigma_2 + \Sigma_3 = 0$ by
%employing the variables
%%
%\begin{subequations}
%\begin{equation}\label{Sigpm}
%\Sigma_1 = -2\Sigma_+\:,\qquad
%\Sigma_2 = \Sigma_+ \mp \sqrt{3} \Sigma_-\:,\qquad
%\Sigma_3 = \Sigma_+ \pm \sqrt{3} \Sigma_-\:.
%\end{equation}
%%
%The connection with the momentum variables $\pi_\pm$
%of~\eqref{pinpi} is
%%
%\begin{equation}\label{Sigpmpi}
%\Sigma_+ = - \frac{\pi_+}{\pi_0} \:,\qquad
%\Sigma_- = -\frac{\pi_-}{\pi_0}\:.
%\end{equation}
%%
%The evolution equations for $\Sigma_\pm$ are
%%
%\begin{equation}
%\Sigma_\pm^\prime = -(2-q)\Sigma_\pm - {}^{3}\!S_\pm\:,
%\end{equation}
%\end{subequations}
%%
%%where  ${}^{3}\!S_+=-\frac{1}{2}{}^{3}\!S_1$ and
%${}^{3}\!S_-=\pm\frac{1}{2\sqrt{3}}({}^{3}\!S_3 -
%{}^{3}\!S_2)$.

%%%%%%%%%%%%%%%%%%%%%%%%%%%%%%%%%%%%%%%%%%%%%%%%%%%%%%%%%%%%%%%%%%%
\section{Scale-automorphism transformations}
\label{Sec:scaleaut}
%%%%%%%%%%%%%%%%%%%%%%%%%%%%%%%%%%%%%%%%%%%%%%%%%%%%%%%%%%%%%%%%%%%
%-----------------------------------------------------------------
%\subsection{The scale group}\label{scale}
%-----------------------------------------------------------------

The \textbf{scale group} is associated with changes of the
length scale. Consider a quantity $\ell$ that has dimension
length, and change the length scale by a constant factor $e^s$:
$\ell \mapsto  e^s\,\ell$. Regarding the metric~\eqref{metric},
it is natural to consider the
1-forms $\hat{\bom}^\alpha$ and the associated structure
constants $\hat{n}_\alpha$ as scale-invariant (i.e., as not
carrying dimension length), which corresponds to viewing the
spatial coordinates as dimensionless. Therefore, $d s^2\mapsto
e^{2s}\,ds^2$ implies
%
%\begin{equation*}
$g_{\alpha\beta} \mapsto e^{2s}g_{\alpha\beta}$,
which leads to
\begin{equation}
\beta^\alpha \mapsto \beta^\alpha +
s\:,\qquad \beta^0 \mapsto \beta^0 + s\:,\qquad
\pi_\alpha \mapsto e^{2s}\,\pi_\alpha\:, \qquad
\pi_0 \mapsto e^{2 s}\,\pi_0\:.
\end{equation}
where the scaling of the canonical momenta is immediate from
the Hamiltonian equations~\eqref{pidef}, since $N \mapsto
e^s\, N$ and thus $\tilde{N} \mapsto e^{-2 s} \tilde{N}$.

\textbf{Spatial frame transformations.}
Consider a linear change of the spatial frame
\begin{subequations}\label{chfr}
\begin{equation}
\hat{\bom}^\alpha  \mapsto A^\alpha{}_\beta\,\hat{\bom}^\beta \:,
\end{equation}
which induces the transformations
\begin{equation}
g_{\alpha\beta} \mapsto
(A^{-1})^\gamma{}_\alpha\,(A^{-1})^\delta{}_\beta\,g_{\gamma\delta}\,,
%\:\:
%C^\alpha{}_{\beta\gamma} \mapsto
%A^\alpha{}_\delta\,(A^{-1})^\phi{}_\beta\,(A^{-1})^\sigma{}_\gamma\,C^\delta{}_{\phi\sigma}\,,\:\:
\qquad
\hat{n}^{\alpha\beta} \mapsto
\frac{1}{\det A}\,A^\alpha{}_\gamma\,A^\beta{}_\delta\,\hat{n}^{\gamma\delta}\,.
\end{equation}
It is of some interest to consider time dependent
transformations, see,
e.g.,~\cite{jan01,jan79,sik80,sik81,chrter06} and references
therein, but for our present purposes it suffices to consider
constant ones. Furthermore, since we consider the diagonal
case~\eqref{diagmetric}, we restrict our attention to diagonal
maps
\begin{equation}\label{scaleaut}
A^\alpha{}_\beta = \diag \big( \exp(a^1), \,\exp(a^2), \,\exp(a^3) \big)\:.
\end{equation}
\end{subequations}
Let
\begin{equation}
a^0 =  \textfrac{1}{3}(a^1 + a^2 + a^3)
\end{equation}
in analogy to $\beta^0$, see~\eqref{pinpi}. Since the
transformation only involves a change of the spatial frame, it
follows that $N \mapsto N$, whence $\tilde{N} \mapsto \exp(3
a^0) \tilde{N}$. Let $(\alpha\beta\gamma)$ be a permutation of
$(123)$; then~\eqref{chfr} and~\eqref{pidef} lead to
\begin{subequations}\label{metricauttransf}
\begin{alignat}{2}
& \beta^\alpha \mapsto \beta^\alpha - a^\alpha\:,\qquad\: & &
\beta^0 \mapsto \beta^0 - a^0\:,\qquad
\hat{n}_\alpha \mapsto \exp(a^\alpha - a^\beta - a^\gamma) \,\hat{n}_\alpha\\
& \pi_\alpha \mapsto e^{-3a^0}\pi_\alpha\:,\qquad\: & &\pi_0
\mapsto  e^{-3a^0}\pi_0 \:.
\end{alignat}
\end{subequations}

\textbf{The group of scale-frame transformations.} The direct
sum of the scale group and the group of (spatial) frame
transformations forms the scale-frame transformations. An
element of this group is represented by a quadruple
$(s,\vector{a}) = (s, a^1,a^2,a^3)$, which acts on the
canonical variables according to
\begin{subequations}\label{scaleautact}
\begin{alignat}{2}
\label{scaleautact1}
& \beta^\alpha \mapsto \beta^\alpha + s - a^\alpha\:,\qquad\:
& & \beta^0 \mapsto \beta^0 + s - a^0\:,\qquad
\hat{n}_\alpha \mapsto \exp(a^\alpha - a^\beta - a^\gamma) \,\hat{n}_\alpha\\
& \pi_\alpha \mapsto e^{2s-3a^0}\pi_\alpha\:,\qquad\:
& &\pi_0 \mapsto  e^{2s-3a^0}\pi_0 \:.
\end{alignat}
\end{subequations}
Furthermore, $\tilde{N} \mapsto \exp(3a^0 -2 s) \tilde{N}$ and
$\hat{n}_\alpha\,e^{2 \beta^\alpha} \mapsto  e^{2 s -3 a^0}\,\hat{n}_\alpha\,e^{2 \beta^\alpha}$
for all $\alpha$ (trivially, if $\hat{n}_\alpha = 0$); note also that $H \mapsto e^{-s} H$.
The energy density $\rho$ is a scalar under~\eqref{chfr} but scales under the scale group;
we obtain
\begin{equation}\label{rho0transf}
\rho\mapsto e^{-2 s} \,\rho \:,\qquad
\rho_0 \mapsto e^{(1+3w)s-3(1+w)a^0}\rho_0 = e^{-(1-w)s} e^{(1+w)(2 s- 3 a^0)} \rho_0 \:.
\end{equation}
From the above it follows that
\begin{equation}\label{TUUact}
T \mapsto e^{2(2 s -3 a^0)}\,T\:,\quad\,
\Ug \mapsto e^{2(2 s -3 a^0)}\,\Ug\:,\quad\, \Uf \mapsto e^{2(2 s -3 a^0)}\,\Uf\:,\quad\,
\mathscr{H} \mapsto e^{2(2 s -3 a^0)}\, \mathscr{H}\:.
\end{equation}

\textbf{The scale-automorphism group.}
The subgroup of spatial frame transformations~\eqref{chfr} that leave
the structure constants invariant is called the (diagonal part of the)
%\textbf{
automorphism
%}
(matrix) group, \textbf{Aut}, of the Lie algebra.
According to~\eqref{scaleautact1}, the automorphism conditions are
\begin{equation}\label{cond}
a^\alpha = a^\beta + a^\gamma \qquad
\forall \alpha \text{ such that } \hat{n}_\alpha \neq 0\:;
\end{equation}
again, $(\alpha\beta\gamma) \in \{(123),(231),(312)\}$.
An alternative representation of the automorphism conditions~\eqref{cond}
%$a^\alpha = a^\beta + a^\gamma$
is $a^\alpha =\textfrac{3}{2}\, a^0$
($\forall \alpha \text{ such that } \hat{n}_\alpha \neq 0$).

The special automorphism group, \textbf{SAut}, is the subgroup of
automorphisms that satisfies $\det A =1$, which corresponds to
$a^0 = 0$. The dimension of SAut is one less than that of Aut.
In Table~\ref{auttable} we give the dimensions of Aut and SAut
for the different Bianchi types.

\begin{table}
\begin{center}
\begin{tabular}{|c|c|c|c|c|c|c|c|c|}
\hline  \multirow{2}{*}{Bianchi type} & \multirow{2}{*}{ScaleFrame} &
\multirow{2}{*}{ScaleAut${}^{\dag}$} &
 \multirow{2}{*}{Aut} &  \multirow{2}{*}{SAut${}^{\ddag}$}
  & \multicolumn{2}{|c|}{Ham.\ scale symm.} & \multicolumn{2}{|c|}{Ham.\ symmetry} \\
& & & &   &vacuum${}^{\dag}$  & fluid  & vacuum & fluid${}^{\ddag}$ \\ \hline
$\mathrm{VIII}$, $\mathrm{IX}$ & \multirow{4}{*}{$4$} & $1$ & $0$ & $0$ & $1$ & $0$ & $0$ & $0$\\
$\mathrm{VI}_0$, $\mathrm{VII}_0$ & & $2$ & $1$& $0$& $2$ & $1$ & $1$ & $0$\\
II &  & $3$ & $2$ & $1$ & $3$ & $2$ & $2$ & $1$\\
I & & $4$ & $3$ & $2$ & $4$ & $3$& $3$ & $2$ \\\hline
\end{tabular}
\caption{This table gives the dimensions of the group of
diagonal scale-frame transformations and its various subgroups:
ScaleAut is the diagonal scale-automorphism group defined
by~\eqref{cond}; Aut (SAut) is the diagonal (special)
automorphism group; the dimension of the group of Hamiltonian
[scale] symmetry transformations, defined in
Subsec.~\ref{HammsymmSec}, depends on if we consider vacuum or
a perfect fluid. The group of Hamiltonian scale symmetry
transformations coincides with ScaleAut in the vacuum case;
likewise, the group of Hamiltonian symmetry transformations
coincides with SAut in the perfect fluid case---this is
indicated by the superscripts ${}^{\dag}$ and ${}^{\ddag}$,
respectively.} \label{auttable}
\end{center}
\end{table}

The direct sum of the scale group and the automorphism group
forms the \textit{scale-automorphism group} \textbf{ScaleAut}.
An element of this group is represented by the quadruple
$(s,\vector{a}) = (s, a^1,a^2,a^3)$, where $(a^1,a^2,a^3)$ is
subject to the automorphism conditions~\eqref{cond}. A
scale-automorphism transformation $(s,\vector{a})$ acts on the
canonical variables according to~\eqref{scaleautact}.
%note that~\eqref{nsaut} holds
%for all $\alpha$ (i.e., for $\alpha$ such that $\hat{n}_\alpha \neq 0$, cf.~\eqref{cond},
%and for $\alpha$ such that $\hat{n}_\alpha = 0$, where the automorphism
%conditions~\eqref{cond} do not apply.)
From~\eqref{TUUact} we see that, in general, the Hamiltonian $\mathscr{H}$
is not invariant under scale-automorphism transformations.

%%%%%%%%%%%%%%%%%%%%%%%%%%%%%%%%%%%%%%%%%%%%%%%%%%%%%%%%%%%%%%%%%%%
\section{ScaleAut and the degrees of freedom}
\label{Sec:scaleautinv}
%%%%%%%%%%%%%%%%%%%%%%%%%%%%%%%%%%%%%%%%%%%%%%%%%%%%%%%%%%%%%%%%%%%

The variables $\Sigma_\alpha$, $N_\beta$ ($\alpha, \beta =1,2,3$)
of~\eqref{sigNdef}
%, i.e.,
%%
%\begin{equation*}
%\Sigma_\alpha = 2 \big( 1 - 3\, \textfrac{\pi_\alpha}{\pi_0} \big)\:,
%\qquad
%N_\alpha = -12 \, \hat{n}_\alpha \, \pi_0^{-1} \, e^{2 \beta^\alpha} \:,
%\end{equation*}
%%
and the time variable $\tau$ of~\eqref{taudef} %, which is given by $d\tau = d
%\beta^0$, cf.~\eqref{taudef},
are invariant under (constant) scale-automorphism
transformations as a direct consequence
of~\eqref{scaleautact}; analogously, $\Omegak$,
${}^3\!S_\alpha$, and $\Omega$ are invariant under ScaleAut. In
the vacuum case, there do not exist any constants on the r.h.\
side of~\eqref{Hdynsys} that are affected by ScaleAut. In the
perfect fluid case with a linear equation of state (where the
constraint $\Omega = 1 - \Sigma^2 - \Omegak$ is used to solve
for $\Omega$) there exists the constant parameter $w$ that
enters~\eqref{Hdynsys} via the deceleration parameter $q$,
see~\eqref{decel}, but $w$ is unaffected by ScaleAut
transformations, as follows from the analysis of
Section~\ref{Sec:scaleaut}. This implies that the reduced
dynamical system~\eqref{Hdynsys} is invariant %(and form-invariant)
under diagonal scale-automorphism transformations.
%The system~\eqref{Hdynsys}
%is a self-contained system of autonomous equations that only
%involve the scale-automorphism invariant variables
%$\Sigma_\alpha$ ($\alpha =1,2,3$), $N_\beta$ (where $\beta$ is
%such that $\hat{n}_\beta \neq 0$), and the scale-automorphism
%invariant constant parameter $w$.

Reconstruction of the metric~\eqref{diagmetric} from a solution
$(\Sigma_\alpha, N_\beta)(\tau)$ of~\eqref{Hdynsys} requires
the Hubble scalar $H$ and an additional `metric' quantity like
$M_\delta$ for each $\delta$ such that $\hat{n}_\delta=0$
(i.e., none in Bianchi type~VIII and~IX, one in
type~$\mathrm{VI}_0/\mathrm{VII}_0$, two in type~II and three
in type~I); see~\eqref{sigNdef} and~\eqref{Mdelta}. However,
these variables are \textit{not} invariant under ScaleAut,
because, by~\eqref{scaleautact},
\begin{equation}
H \mapsto e^{-s}\,H\:,\qquad M_\delta \mapsto
e^{3a^0-2a^\delta}\,M_\delta= e^{a^\beta + a^\gamma -
a^\delta}\,M_\delta \qquad  (\beta\gamma\delta)
\in\big\{(123),(231),(312)\}\:;
\end{equation}
in general, $a^\beta + a^\gamma - a^\delta \neq 0$ (because the
automorphism condition~\eqref{cond} is restricted to $\delta$
such that $\hat{n}_\delta \neq 0$, while, presently, $\hat{n}_\delta =
0$). Therefore, in contrast to $(\Sigma_\alpha, N_\beta)$, the
variables $H$ and $M_\delta$ have `weight' under ScaleAut; hence,
their equations decouple from the
scale-automorphism invariant system~\eqref{Hdynsys} for
`dimensional' reasons.%
\footnote{In~\cite{janugg99} it is shown how one can use a
non-zero inhomogeneous shift vector, determined by the
auto\-morphism group, to construct the metric algebraically and
from a single quadrature for a scale-variable, e.g.,
\mbox{$H=\hat{H}\exp[-\int (1+q)d\tau]$}.}
This decoupling entails that one can obtain the variables $H$
and $M_\delta$ via quadratures from a solution $(\Sigma_\alpha,
N_\beta)(\tau)$ of the system~\eqref{Hdynsys}, i.e.,
\begin{equation}
H=\hat{H}\,\exp\left({-\int} \big(1+q(\tau)\big)\,d\tau\right)\:, \qquad
M_\delta=\hat{M}_\delta\,\exp\left(\int \big(q(\tau)+2\Sigma_\delta(\tau)\big)
\,d\tau\right)\:,
\end{equation}
where $\hat{H}$ and $\hat{M}_\delta$ are constants of
integration. These constants are scale and gauge constants,
respectively, that can be eliminated by means of the
scale-automorphism group: The integration constant $\hat{H}$
can be eliminated by means of a scale-transformation, i.e.,
this integration constant is a scale-parameter. The constants
$\hat{M}_\delta$ can be transformed to $1$ by means of an
appropriate automorphism transformation.

In contrast to the constants $\hat{H}$ and $\hat{M}_\delta$,
the free parameters obtained from solving the reduced
Hubble-normalized system~\eqref{Hdynsys} cannot be eliminated
by means of scale-automorphism transformations, since this
system is invariant under ScaleAut. Consequently, the
system~\eqref{Hdynsys} represents the essential dynamical
content of the present class~A Bianchi cosmologies, and the
variables $(\Sigma_\alpha, N_\beta)$ reflect the degrees of
freedom:%
\footnote{We here define the number of degrees of freedom as
the gauge invariant degrees of freedom, usually called the true
degrees of freedom, minus the scale degree of freedom.}
There are \mbox{$[2 + \text{number of } \hat{n}_\alpha \neq
0]$} degrees of freedom in the perfect fluid case with a linear
equation of state (recall that $\Sigma_1 + \Sigma_2 + \Sigma_3
= 0$), while there are \mbox{$[1 + \text{number of }
\hat{n}_\alpha \neq 0]$} degrees of freedom in the vacuum case
(because of the Gauss constraint).

%The reduced Hubble-normalized system~\eqref{Hdynsys} naturally
%captures the Lie and source contraction hierarchy. By setting
%one of the variables $N_\beta$ to zero in type~VIII or~IX
%(which is equivalent to the Lie algebra contraction
%$\hat{n}_\beta \rightarrow 0$) one obtains precisely the
%scale-automorphism invariant equations for Bianchi type VI$_0$
%or VII$_0$, depending on the relative sign of the remaining two
%$N_\beta$ variables; setting two (one) of the $N_\beta$
%variables to zero in type VIII/IX (VI$_0$/VII$_0$) yields the
%scale-automorphism invariant equations for Bianchi type II;
%setting all three (two; one) $N_\beta$ variables to zero in
%type VIII/IX (VI$_0$/VII$_0$; II) results in the
%scale-automorphism invariant equations for Bianchi type I.
%Setting $\Omega=0$ (which is equivalent to the source
%contraction $\rho_0 \rightarrow 0$) leads to the
%scale-automorphism invariant vacuum equations. Therefore, the
%state spaces associated with the scale-automorphism invariant
%equations for Bianchi types with fewer $\hat{n}_\beta \neq 0$
%form invariant boundary subsets of the state spaces for the
%more general Bianchi types (with a larger number of non-zero
%$\hat{n}_\beta$). Similarly, the scale-automorphism invariant
%vacuum equations appear on an invariant boundary subset of the
%state space associated with the scale-automorphism invariant
%equation for a perfect fluid with a linear equation of state.

When the barotropic equation of state $p = p(\rho)$ is
non-linear, the quantity $w=p/\rho$ is not constant but a
function of $\rho$, i.e., $w = w(\rho)$,
%(A
%simple example is $w = w_0 \rho^\nu$, where $w_0$ and $\nu$ are
%constants, which represents a polytropic equation of state.)
%Note that although $w$ is invariant under ScaleAut (since
%$\rho\mapsto e^{-2 s} \rho$ and $p\mapsto e^{-2 s} p$), the
%function $w(\rho)$ actually changes under ScaleAut (unless
%$w = \mathrm{const}$, which is the case of a linear equation of
%state). We see that,
and thus a degree of freedom is added to the problem;
this can be dealt with
in several ways. First, the system~\eqref{Hdynsys} can be
extended by an evolution equation for $\rho$ (or by the
equation for $H$, since $\rho = 3 H^2 \Omega= 3H^2[1 - \Sigma^2
- \Omegak]$). The thereby enlarged system is not invariant
under ScaleAut but only under the subgroup Aut of ScaleAut.
Alternatively, one can use~\eqref{rhoeq} to express $w$ as a
function of $\beta^0$, i.e., $w = w(\beta^0)$. Instead of
$\rho$ (or $H$) we may thus use $\beta^0$ or a suitable
function of $\beta^0$ as an additional variable;
see~\cite{heietal05}. The enlarged system is invariant under
the subgroup of ScaleAut that is characterized by the condition
$s = a^0$, since $\beta^0 \mapsto \beta^0 + s - a^0$
by~\eqref{scaleautact1}. The relation $d \beta^0 = d \tau$
discloses another possibility: By introducing a
scale-automorphism dependent constant $\hat{\beta}^0$ we can
write $\tau=\beta^0-\hat{\beta}^0$ and regard $w$ as being a
time dependent function $w=w(\tau)$, which turns~\eqref{Hdynsys}
into a non-autonomous system.

In the context of the latter approach, if we assume that the
equation of state is asymptotically linear, i.e., if there
exist $w_{\pm}$ such that $w(\tau) \rightarrow w_{\pm}$ as
$\tau\rightarrow {\pm}\infty$, then by writing $w(\tau)=w_\pm +
f_\pm(\tau)$, where $f_\pm(\tau)\rightarrow 0$ when
$\tau\rightarrow {\pm}\infty$, we can apply a theorem by
Strauss and Yorke~\cite{stryor67} that shows that the future
(past) asymptotics of the non-autonomous system coincide with
the asymptotics of the system with $w=w_+$ ($w=w_-$). Similar
considerations apply to the more general case, where $w(\tau)$
does not converge, but $\liminf_{\tau\rightarrow\pm\infty}
w(\tau)$ and $\limsup_{\tau\rightarrow\pm\infty} w(\tau)$
exist, provided that the asymptotic range of $w(\tau)$ is a
range of structural stability, where the asymptotics of models
are qualitatively similar. This suggests that the case of a
linear equation of state is the cornerstone for any further
asymptotic analysis;
one can use the linear case to
either determine the asymptotic dynamics of the problem when a
limit exists, or to provide bounds for the asymptotic limits
when $\liminf_{\tau\rightarrow\pm\infty} w(\tau)$ and
$\limsup_{\tau\rightarrow\pm\infty} w(\tau)$ exist. These
considerations justify the focus on perfect fluids with linear
equations of state.

%%%%%%%%%%%%%%%%%%%%%%%%%%%%%%%%%%%%%%%%%%%%%%%%%%%%%%%%%%%%%%%%%%%
\section{Hamiltonian structures}
\label{Sec:Hamstruc}
%%%%%%%%%%%%%%%%%%%%%%%%%%%%%%%%%%%%%%%%%%%%%%%%%%%%%%%%%%%%%%%%%%%

This section contains Hamiltonian considerations of a more
general nature: We show how conserved quantities and
monotone functions can be obtained in a rather general context.
In the subsequent section~\ref{Sec:dyncon} we combine these
results with our previous analysis of the scale-automorphism group
and derive conserved
quantities and monotone functions for the reduced dynamical
system~\eqref{Hdynsys}.

Let us consider a Hamiltonian that is of the general form
%
%\begin{equation}\label{Hamqp}
%\Htilde = \nt\,\mathscr{H} = \nt\, \big(T(\vector{p}) + U(\vector{q})\big) =
%\nt\Big(\textfrac{1}{2} G^{ij}\,p_ip_j + U(\vector{q}) \Big) =0\:,
%\end{equation}
%
%
\begin{equation}\label{Hamqp}
\mathscr{H} = T(\vector{p}) + U(\vector{q}) =
\textfrac{1}{2} G^{ij}\,p_ip_j + U(\vector{q}) =0\:,
\end{equation}
where $\{\vector{q},\vector{p}\}$ with $\vector{q} =
(q^i)_{i=0,\ldots,n}$ and $\vector{p} = (p_i)_{i=0,\ldots,n}$
denotes a set of canonical variables.
%$\nt$ is regarded as an independent variable, cf.~\eqref{hameq}.
The kinetic term $T$ is a quadratic form of the momenta; we
assume that $G^{ij}$ is the inverse of a (constant) Lorentzian
metric $G_{ij}$ with signature $(-+\dots +)$. The potential $U$
depends on $\vector{q}$ and may include a number of constants,
collectively denoted by $\kappa$.

Suppose that there is a Lie group of transformations, whose
elements we denote by $(\sigma,\vector{\alpha})$, that acts on
the canonical variables  $\{\vector{q},\vector{p}\}$ and on the
constants $\kappa$ according to
\begin{equation}\label{lietrans}
q^i \mapsto q^i + \sigma - \alpha^i\:, \qquad p_i \mapsto
e^{b\, \sigma + b_j \alpha^j} \, p_i\:, \qquad \kappa \mapsto
e^{d\, \sigma + d_j \alpha^j}\kappa \:,
\end{equation}
where $b\in \mathbb{R}$, $b_j \in \mathbb{R}$ $\forall j$, $d
\in \mathbb{R}$ and $d_j \in \mathbb{R}$ $\forall j$. We assume
that $T$ and $U$ transform identically so that the constraint
$\mathcal{H} = 0$ is preserved. The generator of the
transformation $(\sigma,\vector{\alpha})$ is denoted by $c$ and
its action on an arbitrary function $F$ of the variables
$(\vector{q},\vector{p})$ and the constants $\kappa$ by
$c\!\cdot\!F$. Then~\eqref{lietrans} yields
\begin{subequations}\label{sigalgen}
\begin{align}
\label{sigalgen1}
c\!\cdot\! \vector{q} & = c^i \,\frac{\partial}{\partial q^i}\:\vector{q} =
\Big[ (\sigma-\alpha^0)  \,\frac{\partial}{\partial q^0} +
(\sigma-\alpha^1)  \,\frac{\partial}{\partial q^1} +
\cdots +
(\sigma-\alpha^n)  \,\frac{\partial}{\partial q^n} \Big]\:\vector{q}\:,\\
\label{sigalgen2}
c\!\cdot\!\vector{p} & = \Big( \big[b\, \sigma + b_j
\alpha^j\big] \,p_i\,
\frac{\partial}{\partial p_i}\Big)\:\vector{p} \,,\qquad\quad
c\!\cdot\! \kappa = \Big( \big[d\, \sigma + d_j \alpha^j\big] \, \kappa\,
\frac{\partial}{\partial \kappa} \Big) \: \kappa \:;
%+ \kappa_1(\sigma,\vector{\alpha}) \, \rho_1\, \frac{\partial}{\partial \rho_1}
\end{align}
\end{subequations}
in particular, the action on $\vector{q}$ is represented by a
constant vector $\vector{c} = (c^i)_{i=0,\ldots,n}$ with $c^i =
\sigma - \alpha^i$.

As follows from Noether's theorem, a transformation that leaves
a Hamiltonian $\mathscr{H}$ (form-)invariant, which means that
$\mathscr{H} \mapsto \mathscr{H}$ and that none of the
constants in $\mathscr{H}$ are affected, corresponds to a
variational Hamiltonian symmetry that yields a conserved
momentum quantity.
%see Section~\ref{Sec:Hamicons};
We will refer to such a transformation as a \textit{Hamiltonian
symmetry transformation\/}.

In the present context, $\mathscr{H}$ is given by~\eqref{Hamqp}.
A transformation $(\sigma,\vector{\alpha})$, with generator $c$,
is a Hamiltonian symmetry if $c\!\cdot\!\mathscr{H} = c\!\cdot\! T + c\!\cdot\! U = 0$
and $c\!\cdot\!\kappa = 0$.
The former condition is satisfied if $c\!\cdot\!\vector{p} = 0$
(since $G^{i j}$ is a constant metric); preservation of the constraint ensures that
$c\!\cdot\! U = 0$.
%Expressed in terms of the generator $c$ associated with $(\sigma,\vector{\alpha})$
%we find $c\!\cdot\!\mathscr{H} = c\!\cdot\! T + c\!\cdot\! U = 0$;
%hence $c$ generates a Hamiltonian symmetry
%if $c\!\cdot\!\vector{p} = 0$ (because $G^{i j}$ is a constant
%metric) and if $c\!\cdot\!\kappa = 0$.
Therefore, the conditions are
\begin{equation}\label{bdcond}
b\,\sigma + b_j \alpha^j = 0\quad \text{ and }\quad
d\,\sigma + d_j \alpha^j = 0\:.
\end{equation}
Since $c\!\cdot\!\kappa = 0$, we find
$c\!\cdot\! U = c^i\,\partial_i U$ ($ = 0$).
The Hamiltonian equations
yield
$(c^i p_i)\,\dot{} = -c^i\,\partial_i U = 0$, which implies that
the momentum %generating function
%$p_{\vector{c}} \defeq
quantity $c^i p_i$ associated
with $\vector{c}$, is conserved, i.e.,
\begin{equation}\label{conserved}
c^i p_i = \mathrm{const}\:.
\end{equation}

There exists a more general class of `symmetries' that
do not lead to conserved quantities but to monotone
functions; the analysis is somewhat more involved
and thus deserves special attention.

\subsection{Hamiltonian scale symmetries and monotone functions}
\label{hamscalesymmon}

We say that a transformation~\eqref{lietrans} is a
\textit{Hamiltonian scale symmetry transformation} if
$\mathscr{H}$ is mapped to a multiple of $\mathscr{H}$, i.e.,
$\mathscr{H}\mapsto k\mathscr{H}$ for some $k\in\mathbb{R}$,
where each constant in $\mathscr{H}$ remains unchanged;
in other words, the `conformal
class' $[\mathscr{H}] = \{k \mathscr{H}\,|\, k\in\mathbb{R}\}$
is (form-)invariant under a Hamiltonian scale symmetry
transformation. Note that the group of Hamiltonian symmetry
transformations is a subgroup (of codimension one) of the group
of Hamiltonian scale symmetry transformations. We call a
transformation a proper Hamiltonian scale symmetry
transformation if $k\neq 1$ in $\mathscr{H}\mapsto
k\mathscr{H}$.
%In the vacuum case for types II-IX the scale
%transformation yields such a proper Hamiltonian scale symmetry,
%but as we will see there are others as well.

For a proper Hamiltonian scale symmetry transformation merely the constants
$\kappa$ are invariant, i.e.,
$c\!\cdot\!\kappa = 0$.
Therefore, the Hamiltonian scale symmetries satisfy
\begin{equation}\label{kappaj}
d\,\sigma + d_j \alpha^j = 0\:.
\end{equation}
%
%, while $c\!\cdot\!\vector{p} \neq 0$.
The action of $c$ on $T$ and $U$ is proportional
to $T$ and $U$, respectively, i.e.,
\begin{equation}\label{dirhom}
c\!\cdot\!T = r\, T \quad\text{and}\quad c\!\cdot\!U = c^i\,\partial_i U = r \, U\:,
\end{equation}
%
%and the directional
%derivative of $U$ along $\vector{c}$ is proportional to $U$
%itself, i.e.,
%%
%\begin{equation}\label{dirhom}
%c^i\,\partial_i U = r \, U\:,
%\end{equation}
%%
for some $r=\mathrm{const}$; a %constant
rescaling of
$\vector{c}$ is accompanied by the same rescaling of $r$.
It follows that
\begin{equation}\label{dirhom2}
%\dot{p}_{\vector{c}} =
(c^ip_i)\,\dot{} = -c^i\,\partial_i U = -r\,U\:,
\end{equation}
and thus $c^ip_i$ is monotone if $U$ has a
definite sign.
%(In the context of Bianchi class~A cosmology,
%however, these monotone quantities do not necessarily lead to
%monotone functions on the reduced Hubble-normalized
%system~\eqref{Hdynsys}.)

In addition to $c^ip_i$ we construct a more intricate
monotone quantity. Define
\begin{equation}\label{Mmondef}
M \defeq  %b \,p_{\vector{c}} \,\exp\big(\textfrac{1}{2}\, q_{\vector{k}}\big) =
M_0 \,c^j p_j\,\exp\big(\textfrac{1}{2}\,k_i q^i\big)\:,
\end{equation}
where $\vector{c}$ is associated with the generator of a proper
Hamiltonian scale symmetry (i.e., $r\neq 0$), $M_0$ is a
constant, and $\vector{k} = (k^i)_{i=0,\ldots,n}$ is to be
specified; indices are lowered with $G_{i j}$, i.e., $k_i =
G_{i j} k^j$. Hamilton's equations and the Hamiltonian
constraint $\mathcal{H} = 0$ lead to
\begin{equation}\label{Mdot1}
\dot{M} = \textfrac{1}{2}\, M_0
\left[r\,G^{ij} + c^{(i} k^{j)} \right]p_i p_j\,\exp\big(\textfrac{1}{2}\,k_i q^i\big)\:.
\end{equation}
Accordingly, the question of monotonicity of $M$ is determined
by the properties of the quadratic form $\big(r\,G^{ij} +
c^{(i} k^{j)} \big) p_i p_j$. The causal character of
$\vector{c}$ plays a crucial role.

%-----------------------------------------------------------------
%\subsubsection*{A timelike scale symmetry generator}
%-----------------------------------------------------------------

%\textbf{A timelike scale symmetry generator.}
Let us first consider the case of a \textbf{timelike scale symmetry
generator} $\vector{c}$. W.\ l.\ o.\ g.\ we
assume
\begin{equation}
\vector{c}^2 = c_i c^i = -1 \:,
\end{equation}
which fixes $r$ in~\eqref{dirhom} up to a sign;
%however,
we refer to~\eqref{cnonnorm} et seq.\ for the case
$\vector{c}^2 \neq -1$. The choice
\begin{equation}\label{kchoice}
\vector{k} = r \vector{c} \quad (\leftrightarrow\, k^i = r c^i\, ) \:,
%\leftrightarrow \,k^{j^\prime} = r \delta^{j^\prime}_{\weg 0} )\:,
\end{equation}
leads to $r\,G^{ij} + c^{(i} k^{j)} = r \big( G^{ij} +
c^{i} c^{j} \big)$ being positive or negative semidefinite,
which implies that %the quantity $M$,
\begin{equation}\label{montimelike}
M = %b \,p_{\vector{c}}\, \exp\big(\textfrac{1}{2}\, q_{\vector{k}}\big) =
%b \,p_{\vector{c}}\, \exp\big(\textfrac{r}{2}\, q_{\vector{c}}\big) =
M_0 \,c^j p_j\,\exp\big(\textfrac{r}{2} \,c_i q^i\big)\:,
\end{equation}
is a monotone function, see~\eqref{Mdot1}.

%\begin{remark}
%Since
%%
%\begin{equation*}
%c^i p_i = c^{i^\prime} p_{i^\prime} = p_{0^\prime} \:,
%\qquad
%c_j q^j = c_{j^\prime} q^{j^\prime} =
%\eta_{i^\prime j^\prime} c^{i^\prime} q^{j^\prime} = -q^{0^\prime} \:,
%\end{equation*}
%%
%the function $M$ can be viewed as
%%
%\begin{equation}
%M = M_0 \,  p_{0^\prime} \, \exp\big({-\textfrac{r}{2}}\, q^{0^\prime}\big) \:.
%\end{equation}
%%
%\end{remark}

\begin{remark}
It is of interest to note that there exists a non-linear
canonical point transformation from
%$(q^{i^\prime})_{i^\prime=0,\ldots,n}$ to
$(q^{i})_{i=0,\ldots,n}$ to
$(Q^{i^\prime})_{i^\prime=0,\ldots,n}$ such that
\begin{equation}\lb{timemom}
Q^{0^\prime} = %-2 (M_0  r)^{-1} \exp\big(\textfrac{r}{2}\, q^{0^\prime}\big)=
-2 (M_0 r)^{-1} \exp\big({-\textfrac{r}{2}}\, c_i q^i\big)\:,
\qquad P_{0^\prime} = M\:.
\end{equation}
\end{remark}

For practical reasons it is useful to consider the case of a
timelike scale symmetry generator $\vector{c}$, i.e., $c^i
\partial_i U = r U$, that is not normalized, i.e.,
\begin{equation}\label{cnonnorm}
\vector{c}^2 = c_i c^i < 0 \:.
\end{equation}
This results in a straightforward generalization
of~\eqref{montimelike},
\begin{equation}\label{montimelikenn}
M = %b \,p_{\vector{c}}\, \exp\big({-\textfrac{r}{2}}\, \textfrac{1}{\vector{c}^2}\, q_{\vector{c}}\big) =
M_0 \,c^j p_j\,\exp\big({-\textfrac{r}{2}} \, \textfrac{1}{\vector{c}^2}\,c_i q^i\big)\:.
\end{equation}
Since $r^{-1} \vector{c}$ is invariant under rescalings of
$\vector{c}$, cf.~\eqref{dirhom}, this is true for $r
c_i/\vector{c}^{2}$ as well; therefore,~\eqref{montimelike}
and~\eqref{montimelikenn} define the same function $M$, which,
by construction, is monotone.

%-----------------------------------------------------------------
%\subsubsection*{A null scale symmetry generator}
%-----------------------------------------------------------------

%\textbf{A null scale symmetry generator.}
Next we consider the case of
a \textbf{null scale symmetry generator}
$\vector{c}$, i.e.,
\begin{equation}
\vector{c}^2 = c_i c^i = 0 \:.
\end{equation}
The vector $\vector{c}$ cannot be normalized; however, there
exists a second null vector, $\bar{\vector{c}}$, such that
\begin{equation}\label{barc}
\vector{c}\, \bar{\vector{c}} = c^i \bar{c}_i = {-1}\:.
\end{equation}
The choice
\begin{equation}\label{kchoicenull}
\vector{k} = 2 r \bar{\vector{c}}
\quad (\leftrightarrow\, k^i = 2  r \bar{c}^i\,) \:
% \leftrightarrow \,k^{J^\prime} = 2 r \delta^{J^\prime}_{\weg 2} )\:,
\end{equation}
yields a positive or negative semidefinite form
$r\,G^{ij} + c^{(i} k^{j)}  = r \big( G^{ij} + 2\, c^{(i}
\bar{c}^{j)} \big)$, which implies that
\begin{equation}\label{monnull}
M = %b \,p_{\vector{c}}\, \exp\big(\textfrac{1}{2}\, q_{\vector{k}}\big) =
%b \,p_{\vector{c}}\, \exp\big(r\, q_{\bar{\vector{c}}}\big) =
M_0 \,c^j p_j\,\exp\big(r \,\bar{c}_i q^i\big)\:,
\end{equation}
is monotone, see~\eqref{Mdot1}; $M$ is independent of the
choice of scaling of $\vector{c}$, since $r \bar{\vector{c}}$
is invariant under rescalings of $\vector{c}$ --- note that
$r^{-1} \vector{c}$ is invariant because of~\eqref{dirhom}
and $\vector{c} \bar{\vector{c}}$ is invariant because
of~\eqref{barc}.

\begin{remark}
The null case is the marginal
case; in the case of a spacelike scale symmetry generator
there does not exist any choice of $\vector{k}$ such that the
function $M$ in~\eqref{Mmondef} becomes monotone.
%To see this,
%we consider
%%
%\begin{equation}\label{quadmatr}
%\left(r\,\eta^{i^\prime j^\prime} +
%\delta^{(i^\prime}{}_{\!1}\, k^{j^\prime)} \right)
%p_{i^\prime} p_{j^\prime} \:,
%\end{equation}
%%
%which is the analog of~\eqref{quadformtime} in the spacelike
%case, since, w.l.o.g., $c^{i^\prime} =
%\delta^{i^\prime}{}_{\!1}$ ($i^\prime=0,\ldots,n$) for the
%spacelike generator. It is not difficult to prove that this
%quadratic form is indefinite irrespective of the choice of
%$\vector{k}$, which implies that the function $M$ is not
%monotone; we omit the details.
%%It is not difficult
%%to prove that this quadratic form is indefinite irrespective of
%%the choice of $\vector{k}$:
%%The symmetric matrix in~\eqref{quadmatr}
%% possesses the eigenvalue $r$ with multiplicity $n-3$;
%%in addition, there exist three
%%eigenvalues, $\lambda_1$, $\lambda_2$, $\lambda_3$, that might
%%be different from $r$. Analyzing the first two coefficients of
%%the characteristic polynomial
%%of matrix we see that
%%%
%%\begin{equation*}
%%4 \big( \lambda_1 \lambda_2 + \lambda_1 \lambda_3 + \lambda_2 \lambda_3 \big) =
%% -4 r^2 - (k^{0^\prime})^2 - (k^{2^\prime})^2  - (k^{3^\prime})^2 - \cdots  < 0 \:;
%%\end{equation*}
%%%
%%it follows that the eigenvalues $\lambda_1$, $\lambda_2$,
%%$\lambda_3$ cannot possess the same sign. Therefore, the quadratic
%%form~\eqref{quadmatr} is indefinite and the function $M$ is not
%%monotone.
\end{remark}

%-----------------------------------------------------------------
\subsection{Invariant monotone functions}
\label{invmonfct}
%-----------------------------------------------------------------

The monotone functions $M$ of the type~\eqref{Mmondef} are in
general not invariant under the action~\eqref{lietrans} of the
group of transformations. However, as we will show in the
following, we can exploit the freedom of choosing the vectors
$\vector{c}$, $\vector{k}$, and the constant $M_0$ to remedy
this defect.

%Let $(\sigma,\vector{\alpha})$ be a Hamiltonian scale symmetry
%and $c$ its generator.
%it acts on the space of the variables $\vector{q} = (q^i)_{i=0,\ldots,n}$
%as the operator $d^i \partial_i$, see~\eqref{sigalgen1}, and on the
%momenta $\vector{p} = (p_i)_{i=0,\ldots,n}$ according to~\eqref{sigalgen2}.
%(The constants $\rho_0$, $\rho_1$, \ldots, $\rho_m$ remain unchanged
%under a Hamiltonian scale symmetry, which is because
%$\kappa_j(\sigma,\vector{\alpha}) = 0$ $\forall j =0,\ldots,m$,
%see~\eqref{sigalgen2} and~\eqref{kappaj};
%in other words, $d$ is the null operator on the constants.)
Consider the set of Hamiltonian scale symmetries $(\sigma,\vector{\alpha})$
and the set of its generators $c$.
The vectors $\vector{c} = (c^i)_{i=0,\ldots,n}$
%associated with
%the generators % $c$ of the Hamiltonian scale symmetries
acting
on the space of the variables $\vector{q}$
%cf.~\eqref{sigalgen1},
form a linear subspace; the Hamiltonian symmetries are
a subspace of codimension one. Therefore, each generator
$c^i \partial_i$ can be represented as a linear combination
\begin{equation}
c^i = \cp^i + \cs^i \:,
\end{equation}
where $\cp^i$ is associated with a (fixed) generator
of a proper scale symmetry and $\cs^i$ with
the generator of a Hamiltonian symmetry.
%If $\cp^i$ is timelike (or spacelike), then w.l.o.g.\ we may assume
%$\cp^i$ to be normalized, i.e., $G_{i j} \cp^i \cp^j = \pm 1$.

Since we are considering Hamiltonian scale symmetries,
we observe
\begin{equation}
c\!\cdot\! U = c^i \partial_i U = \cp^i \partial_i U + \cs^i \partial_i U = r U
\end{equation}
and $c\!\cdot\! T = r T$, %see~\eqref{dircyc} and~\eqref{dirhom}. % $r$ depends on the choice of $\cp$.
see~\eqref{dirhom}.
From $c\!\cdot\! T = r T$ it follows
that $c$ acts according to
\begin{equation}
c\!\cdot\!\vector{p} = \Big(\frac{r}{2}\: p_j\, \frac{\partial}{\partial p_j}\Big)\,\vector{p}
\end{equation}
on the momenta. Consequently, the generator $c$ acts on the
monotone function $M$ according to
\begin{equation}
c\!\cdot\! M = \Big( c^i \partial_i + \frac{r}{2}\: p_j\, \frac{\partial}{\partial p_j} \Big) \, M =
\textfrac{1}{2} \, \big(  r +  k_i c^i \big) \, M\:.
\end{equation}
Let us restrict our attention to the case of a timelike scale
symmetry generator $\cp$, i.e., we assume that $G_{i j} \cp^i
\cp^j < 0$; then \mbox{$k^i = {-}(r/\vector{\cp}^2)\, \cp^i$}
and the monotone function $M$ is given by
\begin{equation}\label{Mproper}
M = M_0 \,p_j \cp^j\:\exp\big({-\textfrac{r}{2}} \, \textfrac{1}{\vector{\cp}^2}\,G_{ij}\,\cp^i q^j\big)\:,
\end{equation}
cf.~\eqref{montimelikenn}.
We obtain
\begin{equation}\label{dM}
c \!\cdot\! M = \frac{1}{2}  \Big(  r -   G_{ij} \cp^i c^i \,\frac{r}{\vector{\cp}^2}\Big) \, M
= {-\frac{1}{2}} \,\frac{r}{\vector{\cp}^2}\,\Big( {-\vector{\cp}^2}
+ G_{i j} \cp^i (\cp^j + \cs^j) \Big) \, M
= {-\frac{1}{2}}  \,\frac{r}{\vector{\cp}^2}\, G_{i j} \cp^i  \cs^j\,M \:;
\end{equation}
hence, $M$ is not invariant under the group of Hamiltonian
scale symmetries (unless the group of Hamiltonian symmetries is
trivial). However,~\eqref{dM} suggests that, under certain
circumstances, there exists a canonical choice of $\cp^i$ such
that $M$ is invariant under Hamiltonian scale symmetries.

Assume that (i) the vectors $\vector{c}$ associated with the
generators of Hamiltonian scale symmetries form a timelike
space and that (ii) the vectors $\vector{\cs}$ associated with
the generators of of Hamiltonian symmetries are embedded
therein as a spacelike subspace (which is of codimension one).
Under these conditions it is possible to choose a generator
$\vector{\cp}$ of a proper Hamiltonian scale symmetry that is
orthogonal to the spacelike subspace of Hamiltonian symmetries,
i.e.,
\begin{equation}\label{orthrel}
%G_{i j}  \cp^i \cp^j = -1\qquad\text{and}\qquad
G_{i j} \cp^i \cs^j = 0
\end{equation}
\textit{for all} generators $\vector{\cs}$ of Hamiltonian symmetries.
Consider the monotone quantity $M$ constructed
from $\vector{\cp}$ by~\eqref{Mproper}.
In this case (and only in this case) we obtain % this monotone function $M$ is
invariance under Hamiltonian scale symmetries, i.e.,
\begin{equation}\label{dM0}
c \!\cdot\! M = 0
\end{equation}
for all generators $c$ of Hamiltonian scale symmetries.

Although $M$ is invariant under Hamiltonian scale symmetries,
it is not necessarily invariant under transformations that
affect (some of) the constants $\kappa$. Since it is of
interest to achieve this general invariance we utilize the
freedom of choosing $M_0$. Under a transformation
$(\sigma,\vector{\alpha})$ with generator $\tilde{c}$ that is
not necessarily a Hamiltonian scale symmetry $M$ transforms
according to
\begin{equation}\label{hM}
\tilde{c}\!\cdot\!M/M_0 = \big( \tilde{d}\, \sigma +
\tilde{d}_i \alpha^i\big)  \, M/M_0\:,
\end{equation}
where $\tilde{d}$ and $\tilde{d}_i \in \mathbb{R}$, as follows
from a computation based on~\eqref{sigalgen}. Due
to~\eqref{dM0}, %the expression
$\tilde{d}\, \sigma +
\tilde{d}_i \alpha^i$ vanishes if $(\sigma,\vector{\alpha})$ is
a Hamiltonian scale symmetry, cf.~\eqref{kappaj}. Recall that
$\kappa$ is a collection of constants, i.e., ${}^a\!\kappa$,
and~\eqref{kappaj} a collection of conditions, i.e., ${}^a\!d
\,\sigma + {}^a\!d_i \,\alpha^i = 0$, where $a$ ranges in some
unspecific index set. Consequently, $\tilde{d}\, \sigma +
\tilde{d}_i \alpha^i$ is a linear combination of the the linear
expressions ${}^a\!d \,\sigma + {}^a\!d_i \,\alpha^i$, i.e.,
\begin{equation}\label{dDd}
\tilde{d}\, \sigma + \tilde{d}_i \alpha^i = \sum_a D_a \big({}^a\!d \,\sigma + {}^a\!d_i \,\alpha^i \big)
\:,\qquad
(D_a \in\mathbb{R}\quad\forall a)\:.
\end{equation}
If we choose $M_0$ according to
\begin{equation}\label{binrho}
M_0 = \prod_{a} ({}^a\!\kappa)^{-D_a},
\end{equation}
then~\eqref{sigalgen2} and~\eqref{dDd} yield
\begin{equation*}
\nonumber
\tilde{c}\!\cdot\!M_0  =
\left( \sum_a \big( {}^a\!d\, \sigma + {}^a\!d_i \alpha^i \big) \:{}^a\!\kappa\: \frac{\partial}{\partial \,{}^a\!\kappa} \right)
\prod_{a} ({}^a\!\kappa)^{-D_a}
= % -\Big( \sum_a D_a \big({}^a\!d \,\sigma + {}^a\!d_i \,\alpha^i \big) \Big)  \:\prod_{a} ({}^a\!\kappa)^{-D_a} =
-\big( \tilde{d}\, \sigma + \tilde{d}_i \alpha^i\big) \,M_0\:.
\end{equation*}
%
%\begin{align}
%\nonumber
%\tilde{c}\!\cdot\!M_0 & = \tilde{c}\cdot\!\prod_{a} ({}^a\!\kappa)^{-D_a} =
%\left( \sum_a \big( {}^a\!d\, \sigma + {}^a\!d_i \alpha^i \big) \:{}^a\!\kappa\: \frac{\partial}{\partial \,{}^a\!\kappa} \right)
%\prod_{a} ({}^a\!\kappa)^{-D_a}  \\[0.5ex]
%& = -\Big( \sum_a D_a \big({}^a\!d \,\sigma + {}^a\!d_i \,\alpha^i \big) \Big)  \:\prod_{a} ({}^a\!\kappa)^{-D_a} =
%-\big( \tilde{d}\, \sigma + \tilde{d}_i \alpha^i\big) \,M_0\:.
%\end{align}
Therefore, in combination with~\eqref{hM} we arrive at
\begin{equation}
\tilde{c}\!\cdot\! M = (\tilde{c}\!\cdot\!M_0)\, M/M_0 + M_0\: (\tilde{c}\!\cdot\!M/M_0) = 0\:.
\end{equation}
We conclude that there exists a unique choice of $M_0$ in terms
of the constants ${}^a\!\kappa$, see~\eqref{binrho}, that makes
the function $M$ invariant under the transformation group. (We
assume that the constants transform independently; if this is
not the case, uniqueness does not hold in general.)

%-----------------------------------------------------------------
\subsection{Symmetry breaking and monotone functions}\label{symmbreaking}
%-----------------------------------------------------------------

Consider a Hamiltonian $\mathcal{H}$ with a potential $U =
U(\vector{q})$ that
is a sum of a finite number of terms,
%i.e., $U(\vector{q}) = U_1(\vector{q}) + U_2(\vector{q}) + \cdots$, so that
%
\begin{equation}
%\Htilde = \nt\,
\mathscr{H} =  T(\vector{p}) + U(\vector{q}) =
\textfrac{1}{2} G^{ij}\,p_ip_j + U_1(\vector{q}) + U_2(\vector{q}) + %U_3(\vector{q}) +
\cdots  =0\:.
\end{equation}
In general, each of the potential terms transforms differently
under the transformation group; in addition, each term may (or
may not) include a constant (or several constants) that change
under these transformations.

Consider the Hamiltonian
\begin{subequations}
\begin{equation}\label{H1}
%\Htilde_1 = \nt\,
\mathscr{H}_1 =
\textfrac{1}{2} G^{ij}\,p_ip_j + U_1(\vector{q}) = 0\:.
\end{equation}
Assume that this Hamiltonian is associated with a group of Hamiltonian scale
symmetries and Hamiltonian symmetries.
The introduction of an additional term into the Hamiltonian~\eqref{H1},
i.e.,
\begin{equation}\label{H2}
%\Htilde_2 = \nt\,
\mathscr{H}_2 =
\textfrac{1}{2} G^{ij}\,p_ip_j + U_1(\vector{q}) + U_2(\vector{q}) = 0\:,
\end{equation}
\end{subequations}
breaks the (scale) symmetry group in general and
the dimension of the (scale) symmetry group decreases.
%by the number of constants the additional potential term contains.
(By adding more %and more
potential terms we obtain a hierarchy of Hamiltonian problems
and successive symmetry breaking.) However, under certain
conditions, (scale) symmetry breaking does not affect the
monotonicity properties of functions, i.e., the (scale)
symmetries of a simpler Hamiltonian problem generate functions
that may still be monotone functions for a more complex
Hamiltonian problem.

For definiteness, consider the Hamiltonian~\eqref{H1} and
a Hamiltonian scale symmetry associated with~\eqref{H1}, i.e.,
$\mathcal{H}_1$ ($\leftrightarrow$ \,$T$ and $U_1$) is mapped to a multiple
of $\mathcal{H}_1$, $e^{r_1} \mathcal{H}_1$, while constants remain unchanged.
The generator (of the representation of this transformation on $\vector{q}$-space) is
$\vector{c}$, where
\begin{subequations}
\begin{equation}
c^i \partial_i U_1 = r_1 U_1\:,
\end{equation}
cf.~\eqref{dirhom}.
Now consider~\eqref{H2} and assume that the Hamiltonian scale symmetry of $\mathcal{H}_1$
acts on $U_2$ according to
\begin{equation}
c^i \partial_i U_2 = r_2 U_2 \:,
\end{equation}
\end{subequations}
so that $c^i \partial_i (U_1 +U_2) = r_1 U_1 + r_2 U_2$; we assume that $r_1 \neq r_2$.
(If $r_1 = r_2$, then $\vector{c}$ is a scale symmetry of $\mathcal{H}_2$
and the problem reduces to the familiar problem of Subsection~\ref{hamscalesymmon}.)

We introduce two quantities. The first is defined in analogy
with~\eqref{dirhom2}, i.e., $c^i p_i$, which leads to
\begin{equation}%\label{dirhom4}
%\dot{p}_{\vector{c}} =
(c^ip_i)\,\dot{} =
-c^i\,\partial_i U = -\big( r_1\,U_1 + r_2\,U_2 \big)\:,
\end{equation}
from which it follows that $c^i p_i$ is monotone if $r_1\,U_1 +
r_2\,U_2$ has a definite sign. The second quantity is defined
in analogy with~\eqref{Mmondef}, i.e., $M = M_0 \,c^j
p_j\,\exp\big(\textfrac{1}{2}\,k_i q^i\big)$. Since
$\vector{c}$ represents a Hamiltonian scale symmetry of the
Hamiltonian~\eqref{H1}, the function $M$ is monotone (under the
conditions of Subsection~\ref{hamscalesymmon}) for the
Hamiltonian problem~\eqref{H1}. However, despite the symmetry
breaking induced by the potential $U_2$, the function may still
be monotone for the Hamiltonian problem~\eqref{H2}. By means of
Hamilton's equations and the constraint $\mathcal{H} = 0$ we
get
\begin{equation}\label{Mdot2}
\dot{M} = M_0 \left[ \textfrac{1}{2}
\Big( r_1\,G^{ij} + c^{(i} k^{j)} \Big) p_i p_j\,
+ (r_1-r_2)\, U_2 \right] \,\exp\big(\textfrac{1}{2}\,k_i q^i\big)\:.
\end{equation}
The quadratic form $\big(r_1\,G^{ij} + c^{(i} k^{j)} \big) p_i
p_j$ is identical to the one in~\eqref{Mdot1} where $r
\leftrightarrow r_1$. Therefore, if $(r_1-r_2)\, U_2$ has the
same sign as the quadratic from $\big(r_1\,G^{ij} + c^{(i}
k^{j)} \big) p_i p_j$, then $M$ is a monotone function. We may
proceed in complete analogy with the analysis of
Subsection~\ref{hamscalesymmon}: In the case of a timelike or
null generator $\vector{c}$
%the quadratic form  $\big(\rg\,G^{ij} +
%c^{(i} k^{j)} \big) p_i p_j$ is given by~\eqref{quadtime}; in
%the null case, it is given by~\eqref{quadnull}. Consequently,
%
we obtain
\begin{equation}
M = % b \,p_{\vector{c}}\, \exp\big(\textfrac{1}{2}\, q_{\vector{k}}\big) =
%b \,p_{\vector{c}}\, \exp\big(\textfrac{r_1}{2}\, q_{\vector{c}}\big) =
M_0 \,c^j p_j\,\exp\big({-\textfrac{r_1}{2}}
\textfrac{1}{\vector{c}^2}\,c_i q^i\big)\:, \qquad
M = %b \,p_{\vector{c}}\, \exp\big(\textfrac{1}{2}\, q_{\vector{k}}\big) =
%b \,p_{\vector{c}}\, \exp\big(r_1\, q_{\bar{\vector{c}}}\big) =
M_0 \,c^j p_j\,\exp\big(r_1 \,\bar{c}_i q^i\big)\:,
\end{equation}
respectively; in both cases, the function $M$ is monotone.

%%%%%%%%%%%%%%%%%%%%%%%%%%%%%%%%%%%%%%%%%%%%%%%%%%%%%%%%%%%%%%%%%%%%%%%%%%%%%%%%%
\section{Dynamical consequences of the scale-automorphism group}
\label{Sec:dyncon}
%%%%%%%%%%%%%%%%%%%%%%%%%%%%%%%%%%%%%%%%%%%%%%%%%%%%%%%%%%%%%%%%%%%%%%%%%%%%%%%%%

In Section~\ref{Sec:scaleautinv} we have understood the reduced
dynamical system~\eqref{Hdynsys}, which contains the
`essential' dynamics of Bianchi class~A models, as a
\textit{kinematical\/} consequence of the scale-automorphism
group; in the following we derive the `essential'
\textit{dynamical\/} consequences of the scale-automorphism
group. We apply the results of the previous section to
construct scale-automorphism invariant conserved quantities and
monotone functions. These structures are expressed in terms of
the scale-automorphism invariant state vector of the reduced
dynamical system~\eqref{Hdynsys} and yield restrictions on the
flow of the scale-automorphism invariant reduced dynamical
system~\eqref{Hdynsys}.

The perspective here is to start with the Hamiltonian
representing the simplest class~A model, the vacuum Bianchi
type~I model, characterized by $\hat{n}_1 = \hat{n}_2 =
\hat{n}_3 =0$ and $\rho_0 = 0$ (and thus a zero potential).
Successively, we introduce potential terms including non-zero
constants that lead to a hierarchy of increasingly complex
problems. For the Hamiltonian symmetry and Hamiltonian scale
symmetry groups the introduction of a new constant
\textit{breaks\/} the previous symmetry group and decreases its
dimension by one. This is because an additional non-zero
constant leads to a new constraint on the transformation
$(s,\vector{a})$, cf.~Section~\ref{Sec:scaleaut}. This
structure naturally motivates a case-by-case study of the
hierarchy associated with the constants $\hat{n}_1$,
$\hat{n}_2$, $\hat{n}_3$, and $\rho_0$.

We begin by discussing the group of Hamiltonian symmetry
transformations and Hamiltonian scale symmetry transformations
for our particular problems.

\subsection{ScaleAut and Hamiltonian (scale) symmetries}
\label{HammsymmSec}

%(which we denote by $\rho_0$, $\rho_1$, $\rho_2$, \ldots, $\rho_m$).
The class~A Hamiltonian $\mathscr{H}$ is an example of a
Hamiltonian that possesses the structure~\eqref{Hamqp}, i.e.,
$\mathscr{H} = T + U = T + \Ug + \Uf$, where $T$, $\Ug$, and
$\Uf$ are given in~\eqref{hamspecific}. To be able to apply the
results of Sec.~\ref{Sec:Hamstruc}, we make the following
identifications:
\begin{equation}
q^i \leftrightarrow \beta^\alpha \:,\qquad
p_i \leftrightarrow \pi_\alpha \:,\qquad
G^{i j} \leftrightarrow \textfrac{1}{2}\,\cg^{\alpha\beta} \:,\qquad
G_{i j} \leftrightarrow 2\cg_{\alpha\beta} \:.
\end{equation}
%
%This implies that, e.g., $\vector{\cp}^2 = G_{i j} \cp^i \cp^j$
%becomes $2 \cg_{\alpha\beta} \cp^\alpha \cp^\beta$
%(whereas, e.g., $c^i p_i$ directly corresponds to $\cp^\alpha \pi_\alpha$).
In the present context, the Hamiltonian is $\Htilde = \nt
\mathscr{H}$. Assuming that $\nt$ scales like a power of
$\mathscr{H}$ under the transformation group, which it does for
the choice~\eqref{taudef}, $\mathscr{H}$ and $\Htilde$ share
the same conserved quantities and monotone functions.

Consider a scale-frame transformation $(s,\vector{a})$ as given
by~\eqref{scaleautact}, where $(s,\vector{a})$ takes the place
of $(\sigma,\vector{\alpha})$ of Sec.~\ref{Sec:Hamstruc}.
This transformation is a \textit{Hamiltonian scale symmetry} if
the constants in $\mathscr{H}$ remain unchanged,
cf.~\eqref{kappaj}. These constants are the structure constants
that appears in $\Ug$ and, in the fluid case, the constant
$\rho_0$ in $\Uf$.

Eq.~\eqref{cond} implies that,
in the \textit{vacuum case}, i.e., $\Uf = 0$,
each transformation
\begin{equation}\label{Hammscalsymm}
(s, \vector{a}) \in \mathrm{ScaleAut}
\end{equation}
is a Hamiltonian scale symmetry, see Table~\ref{auttable}. A
proper Hamiltonian scale symmetry transformation satisfies $2 s
- 3 a^0 \neq 0$, cf.~\eqref{TUUact}; in particular, a scale
transformation is automatically a proper Hamiltonian scale
symmetry.
In the \textit{fluid} \textit{case} with a linear equation of state,%
\footnote{The requirement $\Uf \mapsto k \Uf$ ($k \neq 1$)
is compatible only with a linear equation of state.}
$\Uf$ breaks this symmetry. This is due to the presence of
the constant $\rho_0$, which is affected by
ScaleAut according to~\eqref{rho0transf}.
We conclude that
\begin{equation}
\label{a02}
\big\{ (s,\vector{a}) \in \mathrm{ScaleAut} \:\big|\: (1+3w)s-3(1+w)a^0 = 0 \:\big\}
\subset \mathrm{ScaleAut}
\end{equation}
is the group of Hamiltonian scale symmetries in the fluid case.

The group of \textit{Hamiltonian symmetries} is determined by
the additional condition that $\mathscr{H}$ remain unchanged,
cf.~\eqref{bdcond}. This condition is $2 s - 3 a^0 = 2 s - a^1
-a^2 - a^3 = 0$, which follows from~\eqref{TUUact}.

As a consequence, in the \textit{vacuum case}, i.e., $\Uf = 0$,
\begin{equation}\label{Hammsymm}
  %\mathrm{Ham.\ symm.} =
  \big\{ (s,\vector{a}) \in \mathrm{ScaleAut}\:\big|\:
  s = \textfrac{3}{2} \, a^0 = \textfrac{1}{2} \,(
  a^1 + a^2 + a^3)\:\big\} \subset \mathrm{ScaleAut}
\end{equation}
is the group of Hamiltonian symmetry transformations.
In the \textit{fluid case},
the fluid potential $\Uf$ breaks the symmetry~\eqref{Hammsymm}.
We find that
\begin{equation}\label{Hammsymm2}
  %\mathrm{Ham.\ symm.} =
  \mathrm{SAut} =\big\{ (s,\vector{a}) \in \mathrm{ScaleAut}\:\big|\:
  s = a^0 = 0\:\big\}
\end{equation}
is the group of Hamiltonian symmetries;
it is of codimension $2$ within ScaleAut, see Table~\ref{auttable}.

\begin{remark}
Recall that we assume $-1< w <1$, $w \neq -1/3$. When $w = -1$,
the group of Hamiltonian scale symmetries is Aut, while for $w
= -\textfrac{1}{3}$ this group is the direct sum of the scale
group and SAut. In the stiff fluid case $w = 1$, a Hamiltonian
scale symmetry is automatically a Hamiltonian symmetry, and the
group of Hamiltonian symmetries is identical to the one in the
vacuum case. Consequently, the three values $w=-1$,
$w=-\textfrac{1}{3}$, $w=1$ are associated with exceptional
scale-automorphism properties, which in turn lead to extensive
bifurcations. Although it is not difficult to study the
exceptional values, we choose to retain a unified picture and
thus avoid these values.
\end{remark}
%

%As we will see in Section~\ref{Sec:Hamicons},
%While Hamiltonian
%symmetry transformations result in conserved quantities, proper
%Hamiltonian scale symmetry transformations give rise to
%monotone quantities. Eventually both these structures yield
%restrictions for the flow of~\eqref{Hdynsys} on the reduced
%Hubble-normalized state space.

In the following we derive, for each Bianchi type of class A,
conserved quantities and monotone functions based on our
analysis of the (scale) symmetries of the Hamiltonian. The
conserved quantities are linear combinations of momenta, i.e.,
$c^\alpha \pi_\alpha$. Monotone quantities are either linear
combinations of momenta or of the type~\eqref{Mmondef} of
Subsection~\ref{hamscalesymmon}. To obtain scale-automorphism
invariant quantities, which are expressible in terms of state
space variables of the reduced dynamical
system~\eqref{Hdynsys}, we form quotients of conserved/monotone
momentum quantities and/or use the results of
Subsection~\ref{invmonfct} to make $M$ scale-automorphism
invariant.

%-----------------------------------------------------------------
\subsection{Bianchi type I}\label{BIscaleautcomb}
%-----------------------------------------------------------------

The scale-automorphism group ScaleAut coincides with the group
of scale-frame transformations in Bianchi type~I and is thus
four-dimensional; this is because~\eqref{cond} does not imply
any restrictions on $\vector{a}$ in this case. Therefore, every
element $(s,\vector{a}) \in\mathbb{R}\times \mathbb{R}^3$ is an
element of ScaleAut.

In the \textbf{vacuum case}, the group of Hamiltonian symmetry
transformations is given by~\eqref{Hammsymm} and hence
isomorphic to $\mathbb{R}^3$. An element of this group acts on
$\beta^1$, $\beta^2$, $\beta^3$ according
to~\eqref{scaleautact}, i.e., $\beta^1 \mapsto \beta^1 + b^1$,
$\beta^2 \mapsto \beta^2 + b^2$, $\beta^3 \mapsto \beta^3 +
b^3$, where $b^\alpha = \textfrac{1}{2} ({-}a^\alpha + a^\beta
+ a^\gamma)$ is arbitrary; $(\alpha\beta\gamma)$ is a
permutation of $(123)$.
%The generators $\vector{\cs}$ of the Hamiltonian symmetry group
%can be chosen as
%%
%\begin{equation}
%\label{betagenI}
%\frac{\ptl}{\ptl \beta^1}\:,\qquad
%\frac{\ptl}{\ptl \beta^2}\:,\qquad
%\frac{\ptl}{\ptl \beta^3}\:.
%\end{equation}
%%
Accordingly, every vector $\partial/\partial\beta^\alpha$ is a
symmetry generator, and the momenta $\pi_\alpha$, $\alpha
=1,2,3$, and every linear combination thereof (e.g., $\pi_0 =
\pi_1 + \pi_2 + \pi_3$), are conserved; cf.~\eqref{conserved}.
The consequence in terms of the scale-automorphism invariant
variables of the dynamical system~\eqref{Hdynsys}
is that $\Sigma_\alpha =
\mathrm{const}$ $\forall \alpha$, cf.~Eq.~\eqref{sigdef}.
Since $\Sigma_1 + \Sigma_2 +
\Sigma_3 = 0$ and $\Sigma^2 = 1$, cf.~\eqref{constraints}, we
obtain a circle of fixed points of the system~\eqref{evol}, the
Kasner circle, whose
existence can thus be
viewed as a direct consequence of the scale-automorphism group
and its properties.

The group of Hamiltonian scale symmetry transformations ($=
\mathrm{ScaleAut}$, cf.~\eqref{Hammscalsymm}) acts multiply
transitively on the variables $(\beta^1,\beta^2,\beta^3)$,
while the subgroup of Hamiltonian symmetry transformations acts
simply transitively; hence, Hamiltonian scale symmetries do not
lead to additional insights, since the Hamiltonian symmetries
completely determine the dynamics.

In the \textbf{perfect fluid case}, the group of Hamiltonian
symmetries is SAut, cf.~\eqref{Hammsymm2}, which
is two-dimensional in Bianchi type~I; it acts on
$\beta^1$, $\beta^2$, $\beta^3$ according to
\mbox{$\beta^1 \mapsto \beta^1 - a^1$},
$\beta^2 \mapsto \beta^2 - a^2$,
$\beta^3 \mapsto \beta^3 - a^3$,
where $a^1 + a^2 + a^3 = 0$. The generators of
these transformations are
\begin{equation}\label{typeIIscalesautinv}
\cs = {-a^\alpha} \frac{\partial}{\partial\beta^\alpha}
\qquad\: (\text{with } a^1 + a^2 + a^3 = 0)\:,
\end{equation}
which form a spacelike subspace.
%Consequently, a convenient choice of generators of SAut
%is to take any pair of
%%
%\begin{equation}\label{sautgen}
%\frac{\ptl}{\ptl\beta^1} -
%\frac{\ptl}{\ptl\beta^2}\:,\qquad
%\frac{\ptl}{\ptl\beta^2} - \frac{\ptl}{\ptl\beta^3}\:,\qquad
%\frac{\ptl}{\ptl\beta^3} -
%\frac{\ptl}{\ptl\beta^1}\:,
%\end{equation}
%%
%which span a spacelike surface.
Using these generators we obtain
\begin{equation}
\cs^\alpha \pi_\alpha =
\cs^1 \pi_1 + \cs^2 \pi_2 + \cs^3 \pi_3 = \mathrm{const}
\qquad\: (\text{with }\, \cs^1 + \cs^2 + \cs^3 = 0)\:,
\end{equation}
cf.~\eqref{conserved}; in particular we obtain conservation of
$\pi_1 - \pi_2$, $\pi_1 - \pi_3$, and $\pi_2 - \pi_3$.

The group of Hamiltonian scale symmetry transformations is
three-dimensional. A Hamiltonian scale symmetry %$(s, \vector{a})$
acts on the variables $\beta^1$, $\beta^2$, $\beta^3$
according to~\eqref{scaleautact}, i.e.,
$\beta^1 \mapsto \beta^1 + b^1$,
$\beta^2 \mapsto \beta^2 + b^2$,
$\beta^3 \mapsto \beta^3 + b^3$,
where we use~\eqref{a02} to see that
\begin{equation}\label{Itimelike}
b^\alpha = s - a^\alpha = \frac{3(1+w)}{1+3w} \, a^0 - a^\alpha =
%\frac{1+w}{1+3 w}\,(a^1 + a^2 + a^3) - a^\alpha =
\frac{1}{1+ 3 w}\,\Big( {-2} w a^\alpha + (1+w) a^\beta + (1 +w) a^\gamma\Big)
\end{equation}
is arbitrary; $(\alpha\beta\gamma)$ is a permutation of
$(123)$. In accordance with the assumptions~\eqref{orthrel} we
single out a  proper Hamiltonian scale symmetry transformation
whose generator is timelike and orthogonal to the spacelike
surface~\eqref{typeIIscalesautinv}. Hence we take
$(s,\vector{a})$ with $a^\alpha = a^0 \neq 0$ $\forall \alpha$
and $s$ according to~\eqref{a02}. This leads to $\beta^1
\mapsto \beta^1 + b^0$, $\beta^2 \mapsto \beta^2 + b^0$,
$\beta^3 \mapsto \beta^3 + b^0$, where $b^0 = s - a^0 = 2
a^0/(1+ 3 w)$. The infinitesimal generator of this
transformation is
\begin{equation}\label{Ihomgen}
\vector{\cp} = \frac{\ptl}{\ptl \beta^1} +
\frac{\ptl}{\ptl \beta^2} + \frac{\ptl}{\ptl \beta^3}\:.
\end{equation}
%
%Evidently, this generator can be combined with (two of) the
%generators~\eqref{sautgen} of SAut to yield a complete set of
%generators for the Hamiltonian scale symmetry group.

Since $U$ has a sign, we obtain that the conjugate momentum,
$\cp^\alpha \pi_\alpha = \pi_1 + \pi_2 + \pi_3 = \pi_0$, is a
monotone function; cf.~\eqref{dirhom2}. Likewise, $\pi_\alpha$
is monotone for each $\alpha$, which is immediate from the
monotonicity of $\pi_0$ and the conservation of $\pi_\alpha
-\pi_\beta$ ($\alpha,\beta =1,2,3$).
%one way of understanding this is to regard the
%perfect fluid term as a symmetry breaking term of the vacuum
%case, however, below we are also going to explain monotonicites
%in terms of a Hamiltonian scale symmetry).
The implications for the scale-automorphism invariant variables
$\Sigma_\alpha$, $\alpha =1,2,3$, are the
following:~\eqref{sigdef} entails that that $\Sigma_1 \propto
\Sigma_2 \propto \Sigma_3 \propto \pi_0^{-1}$, and hence
$\Sigma_\alpha$ is monotone $\forall \alpha$.

An intimately related result is obtained when we construct a
monotone quantity $M$ of the type~\eqref{Mmondef}. The
Hamiltonian scale symmetry generator~\eqref{Ihomgen} is
timelike whose squared norm w.r.t.\ $G_{\alpha\beta} = 2
\mathcal{G}_{\alpha\beta}$ (so that $G^{\alpha\beta} =
\textfrac{1}{2}\, \mathcal{G}^{\alpha\beta}$) is
$\vector{\cp}^2 = {-12}$. %2 \mathcal{G}_{\alpha\beta} \cp^\alpha \cp^\beta = -12 \:.
Since $U = \Uf$ is given by~\eqref{fluidpot} we obtain
\begin{equation}
\cp^\alpha \frac{\partial}{\partial \beta^\alpha}\, U = r \,U = 3 \,(1-w) \,U\:.
\end{equation}
Consequently, $\cp^\alpha \pi_\alpha =
\pi_0$, and $c_{\mathrm{p}\,\alpha}\, \beta^\alpha = 2
\mathcal{G}_{\alpha\delta} \beta^\alpha \cp^\delta = {-12}
\beta^0$. Insertion into~\eqref{Mproper} yields
\begin{equation}
M =  %b \,\pi_{\vector{c}}\,
%\exp\big({-\textfrac{r}{2}}\, \textfrac{1}{\vector{c}^2}\, \beta_{\vector{c}}\big) =
M_0 \, \pi_0\, \exp\big[ {-\textfrac{3}{2}}\, (1-w) \,\beta^0\big]\:.
\end{equation}
In general, this quantity is not scale-automorphism invariant
%, since
%%
%\begin{equation*}
%\pi_0 \exp\big( {-} \textfrac{3}{2}\, (1-w) \,\beta^0\big) \mapsto
%\exp\big(\textfrac{1}{2}\,(1+ 3 w) s - \textfrac{3}{2}\, (1+w) a^0 \big) \,
%\pi_0 \exp\big( {-} \textfrac{3}{2}\, (1-w) \,\beta^0\big)
%\end{equation*}
%%
under a ScaleAut transformation $(s,\vector{a})$.
However, the results of Section~\ref{invmonfct} allow us
achieve scale-automorphism invariance by a suitable choice of $M_0$,
namely $M_0 = \sqrt{\rho_0}$. The so-constructed scale-automorphism
invariant $M$ is then expressible in terms
of the scale-automorphism invariant state vector, cf.~\eqref{Omegadef}:
\begin{equation}
M = \rho_0^{-1/2} \,\pi_0 \exp\big( {-} \textfrac{3}{2}\, (1-w) \,\beta^0\big)
\propto \Omega^{-1/2} = (1 - \Sigma^2)^{-1/2}\:.
\end{equation}
%in the second step we have used~\eqref{Omegadef}.

%By construction, the quantity $M$ (and thus $\Omega$
%and $\Sigma^2$) is a monotone quantity on the state
%space.
The derived conserved and monotonic quantities
are sufficient to completely describe
the dynamics on the type~I perfect fluid state space.
The property
$\Sigma_1 \propto \Sigma_2 \propto \Sigma_3 \propto \pi_0^{-1}$
implies that there exist integration constants
$\hat{s}^\alpha$, $\alpha =1,2,3$, such that
\begin{equation}
\Sigma_\alpha = \hat{s}_\alpha\, \sqrt{\Sigma^2} \:,\qquad
\big(\hat{s}_1 + \hat{s}_2 + \hat{s}_3=0\:,\quad
\hat{s}^2=\textfrac{1}{6}(\hat{s}_1^2 + \hat{s}_2^2 + \hat{s}_3^2)=1 \big)\:.
\end{equation}
The dynamics of the type~I perfect fluid case is thus
completely determined by ScaleAut.
%CU: The point is not that it is determined but that it is determined by ScaleAut.

%-----------------------------------------------------------------
\subsection{Bianchi type II}\label{BIIscaleautcomb}
%-----------------------------------------------------------------

Without loss of generality we consider the representation
$\hat{n}_1 \neq 0$, $\hat{n}_2 = \hat{n}_3 = 0$ of Bianchi
type~II. The scale-automorphism group ScaleAut is
three-dimensional in Bianchi type~II, since~\eqref{cond}
represents one condition on $\vector{a}$:
\begin{equation}\label{BIIScaleAut}
a^1 = a^2 + a^3 \:.
\end{equation}
We choose to view $a^2$ and $a^3$ as free parameters, which implies that
elements of ScaleAut take the form
\mbox{$(s,\vector{a}) = (s, a^2 + a^3, a^2, a^3)$}.

In the \textbf{vacuum case} the group of Hamiltonian symmetry
transformations is given by~\eqref{Hammsymm}, i.e., $s =
\textfrac{3}{2} \, a^0 =  a^2 + a^3$;
it is isomorphic to $\mathbb{R}^2$. % and thus two-dimensional.
A Hamiltonian symmetry acts on the variables
$\beta^1$, $\beta^2$, $\beta^3$ according
to~\eqref{scaleautact}, which takes the form
$\beta^1 \mapsto \beta^1$,
$\beta^2 \mapsto \beta^2 + a^3$,
$\beta^3 \mapsto \beta^3 + a^2$.
The associated generators $\cs$ span a timelike surface, which we may write
as $\langle\partial/\partial \beta^2, \partial/\partial \beta^3\rangle$.
Accordingly, the
momenta $\pi_2$ and $\pi_3$ are conserved; this leads to
\begin{equation*}
\frac{\pi_2}{\pi_3} = \frac{2-\Sigma_2}{2-\Sigma_3} = \mathrm{const}\:.
\end{equation*}

The group of Hamiltonian scale symmetry transformations
coincides with ScaleAut, cf.~\eqref{Hammscalsymm}. Since $a^1 =
a^2 + a^3$ and thus $a^0 = \textfrac{2}{3} (a^2 +
a^3)$,~\eqref{scaleautact} results in
\begin{subequations}\label{BIIVacHomothet}
\begin{alignat}{2}
& \beta^1 \:\mapsto\: \beta^1 + s - a^2 - a^3  & & =\, \beta^1 + \textfrac{1}{2}\,\big(2 s - 3 a^0 \big)  \:, \\
& \beta^2 \:\mapsto\: \beta^2 + s - a^2  & & =\, \beta^2 + \textfrac{1}{2}\,\big(2 s - 3 a^0 \big) + a^3 \:,\\
& \beta^3 \:\mapsto\: \beta^3 + s - a^3 & & = \,\beta^3 + \textfrac{1}{2}\,\big(2 s - 3 a^0 \big) + a^2\:;
\end{alignat}
\end{subequations}
where $2 s -3 a^0$, $a^2$, and $a^3$ can be viewed as three degrees of
freedom of ScaleAut.
%Note that restriction to the two-dimensional
%subgroup of Hamiltonian symmetry transformations corresponds to
%setting $2 s - 3 a^0 = 0$.
A simple proper Hamiltonian
scale symmetry is %transformation, $2 s - 3 a^0 \neq 0$. The
%simplest choice is
$a^2 = a^3 = s \neq 0$, which yields the generator
$\partial/\partial\beta^1$. Other proper Hamiltonian scale
symmetries are obtained by combining $\partial/\partial\beta^1$
with the Hamiltonian symmetries $\langle\partial/\partial
\beta^2, \partial/\partial \beta^3\rangle$, leading to, e.g., the
scale transformation generator $\partial/\partial \beta^1 +
\partial/\partial \beta^2 +
\partial/\partial \beta^3$ being a proper Hamiltonian scale symmetry
(which is true for all vacuum class~A models.)

The Hamiltonian equation for the momentum
associated with the proper Hamiltonian scale symmetry generated by
$\partial/\partial\beta^1$
is
\begin{equation}\label{HameqBIIvac}
\dot{\pi}_1 = -\tilde{N} \frac{\partial}{\partial \beta^1} \,U = -4 \tilde{N} U\:,
\end{equation}
and hence $\pi_1$ is a monotone function, since $U$ has a sign;
cf.~\eqref{dirhom2}. Likewise, $\pi_0$ is monotone, since
$\pi_0=\pi_1 + \pi_2 + \pi_3$, where
$\pi_2+\pi_3=\mathrm{const}$.

Using the conserved and the monotone momenta we are able to
construct scale-automorphism invariant monotone functions.
Employing~\eqref{sigdefstrich} we obtain that
\begin{equation}
\frac{\pi_1}{\pi_0}= \frac{1}{1 + \pi_1^{-1}(\pi_2 + \pi_3)} \propto 2 - \Sigma_1 \:,\qquad
\frac{\pi_2}{\pi_0} \propto 2 - \Sigma_2\:,\qquad
\frac{\pi_3}{\pi_0} \propto 2 - \Sigma_3
\end{equation}
are monotone, since $\pi_0$, $\pi_1$ are monotone and $\pi_2$,
$\pi_3$ are conserved. Therefore, $\Sigma_1$, $\Sigma_2$,
$\Sigma_3$, are monotone on the (reduced) state space, as are
\begin{equation*}
\frac{\pi_1}{\pi_2} =  \frac{2-\Sigma_1}{2-\Sigma_2}\:,\quad
\frac{\pi_3}{\pi_1} =  \frac{2-\Sigma_3}{2-\Sigma_1}\:,\quad
\frac{\pi_1}{\pi_2+\pi_3} =  \frac{2-\Sigma_1}{4+\Sigma_1}\:.
\end{equation*}
The existence of these conservations and monotonicities provide
sufficient information to obtain the explicit solutions
of~\eqref{Hdynsys} in the Bianchi type~II vacuum case, see,
e.g.,~\cite{heiugg09b}.

\begin{remark}
It is easy to construct monotone quantities $M$ of the
type~\eqref{montimelike} or~\eqref{monnull} from the type~II
scale symmetry group, but none of these can be turned into a
scale-automorphism invariant quantity, since the generators
$\cs$ of the Hamiltonian symmetries form a \textit{timelike}
subspace $\langle\partial/\partial \beta^2,
\partial/\partial \beta^3\rangle$. Therefore, since $M$ cannot
be expressed in terms of the scale-automorphism invariant state
space variables, $M$ does not yield any restrictions on the
dynamics of the system~\eqref{Hdynsys}.
%on the scale- and gauge-invariant
%degrees of freedom; this implies that
%$M$ is useless in connection with the system~\eqref{Hdynsys}.
\end{remark}

In the \textbf{perfect fluid case}, the group of Hamiltonian
symmetry transformations is SAut, cf.~\eqref{Hammsymm2}, which
is one-dimensional in type~II; it acts on $\beta^1$, $\beta^2$,
$\beta^3$ according to $\beta^1 \mapsto \beta^1$, $\beta^2
\mapsto \beta^2 + a^3$, $\beta^3 \mapsto \beta^3 + a^2$, where
$a^2 + a^3 = 0$. 
The generator of this transformation is given by
\begin{equation}\label{IIcyc}
\vector{\cs} = \frac{\ptl}{\ptl\beta^2} - \frac{\ptl}{\ptl\beta^3}\:,
\end{equation}
which is spacelike.
Based on~\eqref{IIcyc} we obtain that $\cs^\alpha \pi_\alpha =
\pi_2 - \pi_3 = \mathrm{const}$ is a conserved momentum.

The momenta $\pi_1$, $\pi_2$, and $\pi_3$ are monotone. To see
this we simply use the Hamiltonian equations
\begin{equation}\label{HamII}
\dot{\pi}_1 = -\tilde{N} \frac{\partial}{\partial \beta^1} \,U =
-\tilde{N} \big( 4 \Ug + (1-w) \Uf \big)\:,\qquad
\dot{\pi}_{2/3} = -\tilde{N} \frac{\partial}{\partial \beta^{2/3}} \,U =
-\tilde{N} (1-w) \Uf \:,
\end{equation}
and note that $\Ug$ and $\Uf$ have the same (positive) sign.
The relations~\eqref{HamII} show that the Bianchi type~II
fluid case can be viewed as arising from symmetry breaking of
either the type~I fluid case or the type~II vacuum case;
cf.~the analysis of Section~\ref{symmbreaking}. Since $\pi_1$,
$\pi_2$, and $\pi_3$ are monotone, $\pi_0 = \pi_1 + \pi_2 +
\pi_3$, and any other positive linear combination, is monotone
as well.

Using the conserved and the monotone momenta we are able to
construct scale-automorphism invariant monotone functions that
are defined in terms of the state space variables; e.g.,
\begin{equation*}
\frac{\pi_2 -\pi_3}{\pi_0} \propto \Sigma_2 - \Sigma_3\:,
\qquad
\frac{\pi_2 -\pi_3}{\pi_2+\pi_3} \propto \frac{\Sigma_2 - \Sigma_3}{4 +\Sigma_1}\:,
\qquad
\frac{\pi_3}{\pi_2} =
1 - \frac{\pi_2 -\pi_3}{\pi_2} \propto \frac{2-\Sigma_3}{2-\Sigma_2}\:.
\end{equation*}

The group of Hamiltonian scale symmetry transformations is
determined by~\eqref{a02} and~\eqref{BIIVacHomothet}. To obtain
a generator of a proper Hamiltonian scale symmetry that is
timelike and orthogonal to~\eqref{IIcyc}, we choose $a^2 = a^3
= \textfrac{3}{4} a^0$. Then the Hamiltonian scale symmetry
condition~\eqref{a02} implies
\begin{equation*}
2s - 3a^0 = 3a^0\,\frac{1 - w}{1 + 3 w}\:,
\end{equation*}
which turns~\eqref{BIIVacHomothet} into
\begin{equation*}
\beta^1 \mapsto \beta^1 + \frac{3}{2}\,\frac{1 -w}{1+3 w}\, a^0\:,\qquad
\beta^2 \mapsto \beta^2 + \frac{3}{4}\,\frac{3 +w}{1+3 w}\, a^0\:,\qquad
\beta^3 \mapsto \beta^3 + \frac{3}{4}\,\frac{3 +w}{1+3 w}\, a^0\:.
\end{equation*}
A convenient generator of this proper Hamiltonian scale symmetry
transformation is
\begin{equation}\label{IIhom}
\vector{\cp} = 2(1-w) \frac{\partial}{\partial \beta^1} + (3+w) \frac{\partial}{\partial \beta^2}
+ (3+w) \frac{\partial}{\partial \beta^3}\:.
\end{equation}
%
%A general element of the Hamiltonian scale symmetry group is
%obtained by linear combinations of $\vector{\cp}$ and $\vector{\cs}$.
To construct a monotone quantity $M$, see~\eqref{Mproper}, we
use this generator $\vector{\cp}$.
%which is a timelike vector and orthogonal to the
%space of Hamiltonian symmetries.
Using~\eqref{dirhom} and computing the squared norm of
$\vector{\cp}$ (w.r.t.\ $G_{\alpha\beta} = 2
\mathcal{G}_{\alpha\beta}$) yields
$\cp^{\alpha} \partial_\alpha  U = r \, U = 8(1-w) \,U$, i.e.,
$r = 8(1-w)$, and $\vector{\cp}^2 = {-4} (3+w) (7-3w)0 < 0$.
Furthermore,
%i.e., $M = b \pi_{\vector{c}}
%\exp\big({-\textfrac{r}{2}}\,\textfrac{1}{\vector{c}^2}\,
%\beta_{\vector{c}} \big)$, where
%
\begin{align*}
%\pi_{\vector{c}} & =
\cp^\alpha \pi_\alpha & = 2 (1-w) \pi_1 + (3 +w) (\pi_2 + \pi_2) \:,   \\
2 \mathcal{G}_{\alpha\gamma} \cp^\alpha\, \beta^\gamma & =
{-4} (3+w) \beta^1 -  2 (5 -w) (\beta^2 + \beta^3)\:.
\end{align*}

The momentum quantity $\cp^\alpha \pi_\alpha$ can be written as
\begin{equation*}
\cp^\alpha \pi_\alpha = \frac{\pi_0}{6}
\Big( 2 (1-w) (2-\Sigma_1) +  (3 +w) (4 -\Sigma_2 - \Sigma_3) \Big)
\propto \pi_0  \big( 16 + (1 + 3 w) \Sigma_1 \big) \:,
\end{equation*}
and
$\exp\big({-\textfrac{r}{2}}\,\textfrac{1}{\vector{\cp}^2}\,
G_{\alpha\gamma} \cp^\alpha \beta^\gamma \big)$ (where
$G_{\alpha\gamma} = 2 \mathcal{G}_{\alpha\gamma}$) becomes
\begin{equation*}
\exp\big({-r}\,\textfrac{1}{\vector{\cp}^2}\,\mathcal{G}_{\alpha\gamma} \cp^\alpha \beta^\gamma\big) =
\exp\Big[ {-2}\,\frac{1-w}{(3+w)(7-3w)}\:\big(2(3+w)\beta^1 + (5-w)(\beta^2 + \beta^3)\big)\Big]\:.
\end{equation*}
Using~\eqref{scaleautact} in connection with the relation $2(a^2 + a^3) = 3
a^0$, cf.~\eqref{BIIScaleAut}, it is straightforward to compute
the behavior of these expressions under a scale-automorphism
transformation; we get
%
%\begin{equation*}
%\cp^\alpha \pi_\alpha \exp\big({-r}\,\textfrac{1}{\vector{\cp}^2}\,\mathcal{G}_{\alpha\gamma} \cp^\alpha \beta^\gamma\big) \,\mapsto\,
%\exp\Big[ \textfrac{2(5-w)}{(3+w)(7-3w)}\, \big((1+3w)s - 3(1+w)a^0 \big)\Big] \,
%\cp^\alpha \pi_\alpha \,\exp\big({-\textfrac{r}{2}}\,\textfrac{1}{\vector{c}^2}\,\beta_{\vector{c}}\big)
%\end{equation*}
%
\begin{equation*}
M/M_0 \,\mapsto \,
\exp\Big[ \textfrac{2(5-w)}{(3+w)(7-3w)}\, \big((1+3w)s - 3(1+w)a^0 \big)\Big] \, M/M_0\:.
\end{equation*}
The comparison with~\eqref{rho0transf} reveals that the choice
of $M_0$ that makes $M$ scale-automorphism invariant is $M_0 =
\rho_0^{-2(5-w)/(3+w)(7-3 w)}$, cf.~\eqref{binrho}. Hence,
\begin{equation}\label{MII}
M = \rho_0^{-2(5-w)/(3+w)(7-3 w)}\,
\cp^\alpha \pi_\alpha
\exp\big({-r}\,\textfrac{1}{\vector{\cp}^2}\,\mathcal{G}_{\alpha\gamma} \cp^\alpha \beta^\gamma\big) \:.
\end{equation}
Using~\eqref{Omegadef} we find
\begin{equation}
M \,\propto\, [16 + (1+3w)\Sigma_1][N_1^{(1-w)(1+3w)}\:\Omega^{2(5-w)}]^\frac{-1}{(3+w)(7-3w)}\:,
\end{equation}
where $\Omega = 1 -\Sigma^2 - N_1^2/12$ according
to~\eqref{constraints}. By construction, the function $M$ is
monotone w.r.t.\ the flow of the dynamical
system~\eqref{Hdynsys}; hence, via the monotonicity principle, see, 
e.g.,~\cite{waiell97,heiugg09a}, we obtain sufficient
information to analyze the global dynamics and asymptotics of
the flow of~\eqref{Hdynsys}. Again, the global dynamics are a
direct consequence of the scale-automorphism group.

%-----------------------------------------------------------------
\subsection{Bianchi types $\bm{\mathrm{VI}_0}$ and $\bm{\mathrm{VII}_0}$}
\label{BVIVIIscaleautcomb}
%-----------------------------------------------------------------

Without loss of generality we consider the representation
$\hat{n}_1 = 0$, $\hat{n}_2 \hat{n}_3 \neq 0$ of Bianchi
types~$\mathrm{VI}_0$ and $\mathrm{VII}_0$. The
scale-automorphism group ScaleAut is two-dimensional in Bianchi
types~$\mathrm{VI}_0/\mathrm{VII}_0$, since~\eqref{cond}
represents two conditions on $\vector{a}$:
\begin{equation}
a^2 = a^1 + a^3 \,,\quad a^3 = a^1 + a^2 \qquad\Rightarrow\qquad
a^1 = 0\,,\quad a^2 = a^3 = \textfrac{3}{2}\, a^0\:.
\end{equation}
We use $a^0$ as the free parameter; hence, each scale-automorphism transformation
is represented by
$(s,\vector{a}) = \big(s,0,\textfrac{3}{2}\,a^0, \textfrac{3}{2}\,a^0\big)$.

In the \textbf{vacuum case}, the group of Hamiltonian symmetry
transformations is one-dimensional and given
by~\eqref{Hammsymm}, i.e., $s = \textfrac{3}{2} \, a^0$. An
element of this group acts on the variables $\beta^1$,
$\beta^2$, $\beta^3$ according to~\eqref{scaleautact}, which
leads to
%\begin{equation}\label{VIVIIcyc}
$\beta^1 \mapsto \beta^1 + \textfrac{3}{2} \,a^0$,
$\beta^2 \mapsto \beta^2$,
$\beta^3 \mapsto \beta^3$.
%\end{equation}
%
The infinitesimal generator is $\cs =
\partial/\partial\beta^1$, which is a null vector.
%\begin{equation}
%\label{betagenVIVII}
%\cs = \frac{\ptl}{\ptl \beta^1}\:;
%\end{equation}
%%
%From~\eqref{betagenVIVII}
%we see that $\beta^1$ is a cyclic variable; hence
We thus obtain conservation of the conjugate momentum $\pi_1$.

The group of Hamiltonian scale symmetries is ScaleAut, cf.~\eqref{Hammscalsymm}, i.e.,
\begin{equation}\label{VIVIIact}
\beta^1 \mapsto \beta^1 + s \:,\qquad
\beta^2 \mapsto \beta^2 + s - \textfrac{3}{2}\, a^0\:,\qquad
\beta^3 \mapsto \beta^3 + s - \textfrac{3}{2}\, a^0\:;
\end{equation}
setting $s = \textfrac{3}{2} \, a^0$ we recover the Hamiltonian
symmetries.
As the generator of a proper
Hamiltonian scale symmetry
%
%\begin{equation}
%\beta^1 \mapsto \beta^1 \:,\qquad
%\beta^2 \mapsto \beta^2 - \textfrac{3}{2}\, a^0\:,\qquad
%\beta^3 \mapsto \beta^3 - \textfrac{3}{2}\, a^0\:;
%\end{equation}
we choose
\begin{equation}\label{VIVIIgen}
\frac{\partial}{\partial \beta^2} + \frac{\partial}{\partial \beta^3}\:.
\end{equation}
Other proper Hamiltonian scale symmetries are obtained by
combining~\eqref{VIVIIgen} with Hamiltonian symmetries; e.g.,
the scale transformation generator $\partial/\partial \beta^1 +
\partial/\partial \beta^2 +
\partial/\partial \beta^3$.

The momentum quantity associated with~\eqref{VIVIIgen} is $\pi_2 + \pi_3$; it is
monotone, since $U$ has a sign; cf.~\eqref{dirhom2}.
Consequently, $\pi_0 = \pi_1 + \pi_2 + \pi_3$ is monotone as
well. We conclude that
\begin{equation}
\frac{\pi_1}{\pi_0}\propto 2 - \Sigma_1 \:,\qquad
\frac{\pi_1}{\pi_2 + \pi_3} \propto \frac{2- \Sigma_1}{4 + \Sigma_1}
\end{equation}
are scale-automorphism invariant monotone quantities (as is $\Sigma_1$ itself).

To construct a monotone quantity $M$ of the
type~\eqref{Mmondef}, we first note that the generator $\cs =
\partial/\partial \beta^1$ spanning the Hamiltonian scale
symmetries is null. By the results of
Subsection~\ref{invmonfct} this makes it impossible to
construct a scale-automorphism invariant quantity $M$ of the
kind~\eqref{Mproper} with a Hamiltonian scale symmetry
generator that is timelike. We thus resort to the case of a
Hamiltonian scale symmetry generator that is null.

The generators of proper Hamiltonian scale symmetries are
%
%\begin{equation}
$\lambda \,\partial/\partial \beta^1 + \partial/\partial \beta^2 + \partial/\partial \beta^3$
with $\lambda \in \mathbb{R}$. There exists a unique choice of
$\lambda$ that yields a null vector: $\lambda = -1/2$.
Therefore, we get
\begin{equation}
\vector{c} = ({-\textfrac{1}{2}},1,1)^{\mathrm{T}}\:,\qquad \vector{c}^2 = 0\:,\qquad
c^\alpha \frac{\partial}{\partial \beta^\alpha}\, U = r \,U = 4\,U \:.
\end{equation}
In the case of a null generator,
the monotone function $M$ is~\eqref{monnull}, i.e., $M = M_0 \,c^\alpha
\pi_\alpha\,\exp\big(r \,\bar{c}_\alpha \beta^\alpha\big)$,
where $\bar{\vector{c}}$ is a complementary null vector, i.e.,
$G_{\alpha\beta} c^\alpha \bar{c}^\beta = 2 \mathcal{G}_{\alpha\beta} c^\alpha \bar{c}^\beta = {-1}$.
We choose $\bar{\vector{c}}$ to be
\begin{equation}\label{barcchoice}
\bar{\vector{c}} = \big(\textfrac{1}{4},0,0\big)^{\mathrm{T}} \:,
\end{equation}
which generates a Hamiltonian symmetry.
This results in
$\bar{c}_\alpha \beta^\alpha =
2 \mathcal{G}_{\alpha\gamma} \bar{c}^\alpha \beta^\gamma = {-\textfrac{1}{2}} (\beta^2 + \beta^3)$,
and accordingly, the monotone quantity $M$ reads
\begin{equation}
M = M_0\,\big({-\textfrac{1}{2}}\,\pi_1 + \pi_2 + \pi_3 \big)\,\exp\big[{-2}\beta^2 - 2\beta^3\big]\:.
\end{equation}
%
%which transforms as
%
%\begin{equation}
%$(M/M_0) \mapsto \exp\big[ {-}(2 s - 3a^0)\big]\: (M/M_0)$
%\end{equation}
%
Under scale-auto\-morphism transformations we 
have $(M/M_0) \mapsto \exp\big[ {-}(2 s - 3a^0)\big]\: (M/M_0)$;
scale-automorphism
invariance is thus achieved by choosing $M_0 = \pi_1 =
\mathrm{const}$. Therefore,
\begin{equation}
M = \pi_1 \,\big( {-\textfrac{1}{2}}\,\pi_1 + \pi_2 + \pi_3 \big)\,\exp\big[{-2}(\beta^2 + \beta^3)\big]
\end{equation}
is a monotone quantity that can be expressed in terms of the
scale-automorphism invariant variables of the reduced dynamical
system~\eqref{Hdynsys}. Using~\eqref{Ndef}
and~\eqref{sigdefstrich} we obtain
\begin{equation}
M \propto (2-\Sigma_1)
\Big( {-}\textfrac{1}{12}(2-\Sigma_1) +  \textfrac{1}{6}(2-\Sigma_2) + \textfrac{1}{6}(2-\Sigma_2) \Big)
(N_2 N_3)^{-1} \propto \frac{(2-\Sigma_1)(2 + \Sigma_1)}{N_2 N_3} \:.
\end{equation}
Using the Gauss constraint
%in the form
%
%\begin{equation}
%1 - \textfrac{1}{4}\Sigma_1^2 - \textfrac{1}{12}(\Sigma_2 - \Sigma_3)^2 -
%\textfrac{1}{12} (N_2 - N_3)^2  = 0\:,
%\end{equation}
%
we find an alternative representation as
\begin{equation}
M \propto \frac{3(4 - \Sigma_1^2)}{N_2 N_3} =
\frac{(\Sigma_2 - \Sigma_3)^2 + (N_2 - N_3)^2}{N_2 N_3}\:.
\end{equation}
This function is a monotone function on the state space
of~\eqref{Hdynsys} and provides detailed information on the
global dynamics, see~\cite{heiugg09a}.

In the \textbf{perfect fluid case}, there do not exist
Hamiltonian symmetries~\eqref{Hammsymm2}, since SAut is
trivial. The fluid potential breaks the symmetry of the
Hamiltonian, for all ScaleAut transformations.

Since the group of Hamiltonian symmetries is trivial, there do
not exist conserved momenta. However, we can regard the perfect
fluid term as breaking the Hamiltonian symmetries and
Hamiltonian scale symmetries of the vacuum type
$\mathrm{VI}_0$/$\mathrm{VII}_0$ case. Hence consider
$(1,0,0)^{\mathrm{T}}$ and $(0,1,1)^{\mathrm{T}}$, cf.~the
generators $\cs$ and~\eqref{VIVIIgen}. Since both $\Ug$ and
$\Uf$ are positive, we obtain the monotone momentum quantities
$\pi_1$ and $\pi_2 + \pi_3$, see Section~\ref{symmbreaking};
note that
\begin{equation}\label{HamVII}
\frac{\partial}{\partial \beta^1}\,U = (1-w)\Uf \,,\qquad\quad
\Big(\frac{\partial}{\partial \beta^2} + \frac{\partial}{\partial\beta^3}\Big)\,U
= 4\Ug + 2(1-w)\Uf \:.
\end{equation}
%
%\begin{subequations}\label{HamVII}
%\begin{align}
%\frac{\partial}{\partial \beta^1}\,U &= (1-w)\Uf \,,\\[0.5ex]
%\Big(\frac{\partial}{\partial \beta^2} + \frac{\partial}{\partial\beta^3}\Big)\,U
%&= 4\Ug + 2(1-w)\Uf \:.
%\end{align}
%\end{subequations}
%
Evidently, $\pi_0$ % = \pi_1 + \pi_2 + \pi_3$ 
and $-2 \pi_1 +
\pi_2 + \pi_3$ are monotone as well. However, since there does not exist
any conserved momentum we cannot construct any
scale-automorphism invariant monotone functions.

The group of Hamiltonian scale symmetry transformations is
one-dimensional in the perfect fluid case. The group acts according
to~\eqref{VIVIIact} where the Hamiltonian scale symmetry
condition~\eqref{a02} must be satisfied. A straightforward
calculation yields
\begin{equation}\label{VIVIIactfl}
\beta^1 \mapsto \beta^1 + \frac{3 (1+w)}{1 + 3 w}\, a^0 \:,\quad
\beta^2 \mapsto \beta^2 + \frac{1}{2}\,\frac{3(1-w)}{1+3 w}\, a^0\:,\quad
\beta^3 \mapsto \beta^3 + \frac{1}{2}\,\frac{3(1-w)}{1+3 w}\, a^0\:.
\end{equation}
We choose to take
\begin{equation}
\vector{\cp} = 2 (1 + w) \frac{\partial}{\partial \beta^1} +
(1-w) \frac{\partial}{\partial \beta^2}
+ (1-w) \frac{\partial}{\partial \beta^3}\:.
\end{equation}
as the generator of this proper Hamiltonian scale symmetry
transformation and construct the associated monotone
quantity of the type~\eqref{Mmondef}; we have
\begin{equation}\label{e0VII}
\vector{\cp} = (2(1+w),1-w,1-w)^\mathrm{T}:\quad
\cp^{\alpha} \frac{\partial}{\partial \beta^\alpha}\, U = r \, U = 4(1-w) \,U\:,\quad\,
\vector{\cp}^2 = {-4} (1-w) (5+3w)\:;
\end{equation}
in particular, $\vector{\cp}$ is a timelike vector.
The monotone quantity is given by~\eqref{Mproper}, where
\begin{align*}
\cp^\alpha \pi_\alpha  & = 2(1+w) \pi_1 + (1-w)(\pi_2 + \pi_2)\:,\\
2 \mathcal{G}_{\alpha\gamma} \cp^\alpha \beta^\gamma  & = {-4}(1-w)\beta^1 - 2(3+w)(\beta^2 + \beta^3)\:,
\end{align*}
and $r$ and $\vector{\cp}^2$ are given by~\eqref{e0VII}. The
quantities $\cp^\alpha \pi_\alpha$ and
$\exp\big({-r}\,\textfrac{1}{\vector{\cp}^2}\,
\mathcal{G}_{\alpha\gamma} \cp^\alpha \beta^\gamma \big)$ can
be written as
\begin{align*}
& \cp^\alpha \pi_\alpha  = \textfrac{1}{6}\pi_0\Big(2(1+w)(2-\Sigma_1) +  (1-w)(4 -\Sigma_2 - \Sigma_3)\Big)
\propto \pi_0 \big(8 - (1+3w) \Sigma_1 \big)\:, \\
& \exp\big({-r}\,\textfrac{1}{\vector{\cp}^2}\, \mathcal{G}_{\alpha\gamma} \cp^\alpha \beta^\gamma \big) =
\exp\Big[ {-\frac{1}{5+3w}}\:\big(2(1-w)\beta^1 + (3+w)(\beta^2 + \beta^3)\big)\Big]\:,
\end{align*}
respectively. Using~\eqref{scaleautact} (which corresponds
to~\eqref{VIVIIact} in the type $\mathrm{VI}_0/\mathrm{VII}_0$
case) it is straightforward to compute the behavior of these
expressions under a scale-automorphism transformation,
\begin{equation*}
%\pi_{\vector{c}}\exp\big({-\textfrac{r}{2}}\,\textfrac{1}{\vector{c}^2}\,\beta_{\vector{c}}\big)
M/M_0
\,\mapsto\,
\exp\Big[ \textfrac{2}{5+3w}\,\big((1+3w)s - 3(1+w)a^0\big)\Big]\,
%\pi_{\vector{c}}\exp\big({-\textfrac{r}{2}}\,\textfrac{1}{\vector{c}^2}\,\beta_{\vector{c}}\big)\:.
M/M_0\:.
\end{equation*}
By the results of Subsection~\eqref{invmonfct} there must exist a choice
of $M_0$ in terms of $\rho_0$ that makes $M$ scale-automorphism
invariant. We find that
\begin{equation}
M = \rho_0^{-2/(5+3w)}\, \cp^\alpha \pi_\alpha
\exp\big({-r}\,\textfrac{1}{\vector{\cp}^2}\, \mathcal{G}_{\alpha\gamma} \cp^\alpha \beta^\gamma \big)
\end{equation}
is the desired quantity; in combination with~\eqref{Omegadef} this
leads to
\begin{equation}
M \,\propto\,
\big[8 - (1+3w)\Sigma_1 \big][(N_2 N_3)^{\frac{1}{4}(1+3w)}\,\:\Omega]^\frac{-2}{5+3w}\:,
\end{equation}
with $\Omega = 1 -\Sigma^2 - (N_2 -
N_3)^2/12$, cf.~\eqref{constraints}. By construction, the function $M$ is monotone
w.r.t.\ the flow of~\eqref{Hdynsys};
via the monotonicity principle we obtain 
sufficient information to analyze the global dynamics and
asymptotics of this flow --- a
consequence of the scale-automorphism group.

%-----------------------------------------------------------------
\subsection{Bianchi types VIII and IX}
\label{BVIIIIXscaleautcomb}
%-----------------------------------------------------------------

For Bianchi types~VIII and~IX the (diagonal) automorphism group
Aut is trivial because of~\eqref{cond}; thus ScaleAut is
one-dimensional and coincides with the scale group.
Therefore, in the \textbf{vacuum case}, each
Hamiltonian scale symmetry transformation is a scale
transformation, $\beta^1 \mapsto \beta^1 + s$,
$\beta^2 \mapsto \beta^2 + s$,
$\beta^3 \mapsto \beta^3 + s$,
with the generator
\begin{equation}\label{VIIIIXgen}
\cp = \frac{\ptl}{\ptl \beta^1} + \frac{\ptl}{\ptl \beta^2} + \frac{\ptl}{\ptl \beta^3}\:,
\end{equation}
which is a timelike vector with norm
${-12}$ w.r.t.\ $G_{\alpha\beta} = 2
\mathcal{G}_{\alpha\beta}$;
we have
\begin{equation}
\cp^\alpha \frac{\partial}{\partial \beta^\alpha}\, U = r\, U =  4\, U \:.
\end{equation}
Consequently, $\cp^\alpha \pi_\alpha = \pi_0$, and $2
\mathcal{G}_{\alpha\gamma} \cp^\alpha \beta^\gamma = {-12}
\beta^0$. Insertion into~\eqref{Mproper} yields a monotone
quantity,
\begin{equation}
M = M_0 \,\cp^\alpha \pi_{\alpha} \exp\big( {-r} \mathcal{G}_{\alpha\gamma} \cp^\alpha \beta^\gamma \big)
= M_0 \,\pi_0 \exp\big( {-2}\,\beta^0\big)\:.
\end{equation}
It is immediate from~\eqref{scaleautact} that $M$ is
scale-automorphism invariant for any (scale-automorphism
invariant) constant $M_0$ (since $\vector{a} = 0$ and thus $a^0 =
0$ for Bianchi type~VIII and~IX). Expressing $M$ in terms of
the state space variables we obtain
\begin{equation}
M \,\propto \,\big|N_1 N_2 N_3 \big|^{\frac{1}{3}}\:,
\end{equation}
and hence $N_1N_2N_3$ is monotone.

In the \textbf{perfect fluid case}, the group of Hamiltonian
scale symmetries is empty. However, we can regard the perfect
fluid term as breaking the scale symmetry of the type~VIII/IX vacuum case;
alternatively, we view the gravitational potential as breaking the
scale symmetry of the type~I perfect fluid case,
cf.~Section~\ref{symmbreaking}.
With $\vector{\cp}$ as in~\eqref{VIIIIXgen} we obtain
\begin{equation}
\cp^\alpha \frac{\partial}{\partial \beta^\alpha}\, U = \rg\, \Ug + \rf \,\Uf =
4\, \Ug + 3 (1- w) \,\Uf \:,
\end{equation}
cf.~\eqref{Mdot2}. We note that $\Uf > 0$,
cf.~\eqref{fluidpot}, and $0 <\rf < \rg$, when
$-\textfrac{1}{3} < w < 1$. In the case ${-1} < w < -\textfrac{1}{3}$
we observe $0 < \rg < \rf$.
Therefore, the assumptions of
Section~\ref{symmbreaking} are satisfied, and the quantity
\begin{equation}
M \,\propto\, \cp^\alpha \pi_{\alpha}
\exp\big(  {-r} \mathcal{G}_{\alpha\gamma} \cp^\alpha \beta^\gamma\big)\,\propto\,\big|N_1 N_2 N_3 \big|^{\frac{1}{3}}
\end{equation}
is monotone, as is $N_1N_2N_3$, like in the vacuum
case.

%%%%%%%%%%%%%%%%%%%%%%%%%%%%%%%%%%%%%%%%%%%%%%%%%%%%%%%%%%%%%%%%%%%
\section{Discussion}
\label{discussion}
%%%%%%%%%%%%%%%%%%%%%%%%%%%%%%%%%%%%%%%%%%%%%%%%%%%%%%%%%%%%%%%%%%%

In this paper we have analyzed the diagonal scale-automorphism
group of the diagonal vacuum and orthogonal perfect fluid
class~A Bianchi models and its kinematical and dynamical
consequences. 
The main kinematical consequence of the scale-automorphism group ScaleAut 
is the existence of a reduced system of equations.
ScaleAut makes it possible to
reduce the number of coupled equations and to 
construct a reduced system whose dimension is determined by ScaleAut;
equivalently, it is possible,
at least in principle, to produce a single ODE 
whose order is determined by ScaleAut (for results in this
area see, e.g.,~\cite{chrter06,chrter07,terchr09}). 
However, it is
only the most special Bianchi models, %those that are associated 
with the largest automorphism groups, 
that admit sufficiently many conserved quantities to
allow the construction of explicit solutions.

Of particular importance is the hierarchy 
of reduced dynamical systems and accompanying structures that 
are induced by the scale-automorphism group; it is this hierarchy that
makes it possible to qualitatively analyze the dynamics of all models.
As a dynamical consequence, the
scale-automorphism group induces a hierarchy of conserved
quantities and monotone functions in class~A vacuum and perfect
fluid models (with $w=\mathrm{const}$) as given in
Tables~\ref{constable} and~\ref{montable}, respectively.
%These results completely
%exhaust the dynamical consequences of the (continuous)
%scale-automorphism group. Further progress in the context of
%the dynamics of (vacuum and orthogonal perfect fluid) Bianchi
%models must therefore rely on different methods, which are not
%directly related to physical first principles.

\begin{table}
\begin{center}
\begin{tabular}{|cc|c|} \hline
Bianchi type & Matter & Conserved quantities \\ \hline &  & \\[-2.2ex]
I & vacuum & $\Sigma_1$, $\Sigma_2$, $\Sigma_3$, where $\Sigma^2 = 1$ \\
I & perfect fluid & $\Sigma_1/\Sigma_2$, $\Sigma_2/\Sigma_3$, $\Sigma_3/\Sigma_1$ \\
II ({\small $\hat{n}_1 \neq 0$}) & vacuum & $(2-\Sigma_2)/(2-\Sigma_3)$  \\\hline
\end{tabular}
\end{center}
\caption{Conserved quantities for the reduced dynamical system of class~A Bianchi models.}
\label{constable}
\end{table}

\begin{table}
\begin{center}
\begin{tabular}{|cc|c|} \hline
Bianchi type & Matter & Monotone quantities \\ \hline &  & \\[-2.2ex]
I & fluid & $\Sigma^2$ \\[0.5ex]
II % ({\small $\hat{n}_1 \neq 0$})
& vacuum &
$\Sigma_1$, $\Sigma_2$, $\Sigma_3$,  \\
& &   $(2-\Sigma_1)/(2 - \Sigma_2)$,  $(2-\Sigma_3)/(2 - \Sigma_1)$, $(2-\Sigma_1)/(4 +
    \Sigma_1)$ \\[0.5ex]
II % ({\small $\hat{n}_1 \neq 0$})
& fluid &
$\Sigma_2 - \Sigma_3$, $(\Sigma_2-\Sigma_3)/(4 + \Sigma_1)$,
    $(2-\Sigma_2)/(2 - \Sigma_3)$ \\
& & $[16 + (1+3 w)\Sigma_1]\:[N_1^{(1-w)(1+3 w)}\:\Omega^{2(5-w)}]^{-1/(3+w)(7-3w)}$ \\[0.5ex]
$\mathrm{VI}_0$/$\mathrm{VII}_0$ %({\small $\hat{n}_2\hat{n}_3\neq 0$})
& vacuum &
 $\Sigma_1$,
    $(2 - \Sigma_1)/(4 + \Sigma_1)$, \\
& & $(4
    -\Sigma_1^2)/N_2 N_3 = [(\Sigma_2 - \Sigma_3)^2 + (N_2
    - N_3)^2]/N_2 N_3$  \\[0.5ex]
$\mathrm{VI}_0$/$\mathrm{VII}_0$% ({\small $\hat{n}_2\hat{n}_3\neq 0$})
& fluid &
 $[8 -(1+3w)\Sigma_1][(N_2N_3)^{(1+3w)/4}\:\Omega]^{-2/(5+3w)}$  \\[0.5ex]
 VIII/IX & vacuum &     $N_1N_2N_3$  \\[0.5ex]
VIII/IX & fluid &    $N_1N_2N_3$ \\\hline
\end{tabular}
\end{center}
\caption{Monotonic quantities for the reduced dynamical system of class~A Bianchi models.
For Bianchi type~II we assume $\hat{n}_1 \neq 0$; moreover,  $\Omega = 1 -\Sigma^2 - N_1^2/12$.
For Bianchi types $\mathrm{VI}_0$/$\mathrm{VII}_0$ we assume $\hat{n}_2\hat{n}_3\neq 0$;
here, $\Omega = 1 -\Sigma^2 - (N_2 -N_3)^2/12$.}
\label{montable}
\end{table}

In this paper, we have derived
the structures that 
are necessary to describe the dynamics of Bianchi class~A models
from first principles.
Note, however, that the Hamiltonian techniques we have employed are
merely a convenient tool for an intermediate step; our final
results are described in terms of scale-automorphism invariant
Hubble-normalized reduced state vector, which is independent of
a Hamiltonian formulation. Our results exclusively rely on the
scale-automorphism group and could in principle have been
derived without any reference to a symplectic structure. This
does not mean that the Hamiltonian methods and results are not
of interest. In~\cite{ashetal93,ashetal93b} conserved
quantities played a key role for quantizing various spatially
homogeneous models. However, we have shown that, in a classical
context, monotone functions are at least as important as
conserved quantities. Should this not be reflected in a quantum
context as well?  And if so, how? Note that, e.g., monotone
functions associated with timelike generators can be regarded
as timelike momenta, cf.~\eqref{timemom}. If such a momentum
had been conserved instead of monotone, this would have
resulted in a natural frequency decomposition.

The present vacuum and perfect fluid models serve as an example
that illustrates a general mechanism. Instead we could have
considered, e.g., electromagnetic fields or collisionless
(Vlasov) matter. Einstein-Vlasov shows that the
scale-automorphism group has consequences on two levels. First,
it has direct consequences for integrating the matter
equations. In the Vlasov case this means that there is a
connection between the scale-automorphism group and the
conservation of momenta of the particles of the collisionless
gas. Conservation of momenta subsequently plays a key role for
the solution of the Vlasov equation, which in turn is important
for the description of the source. Second, the
scale-automorphism group plays a similar role as in the present
case. It leads to a hierarchy of structures associated with a
hierarchy of distribution functions, including distributional
ones. It is possible to perform a similar analysis as in the
present case and tie the key structures that have entered into
the theorems about spatially homogeneous Einstein-Vlasov
systems, see~\cite{rentod99,renugg00,heiugg06}, to the
scale-automorphism group.

The automorphism group is what remains of the spatial
diffeomorphism group in the context of Bianchi
symmetries~\cite{jan79}. The spatial diffeomorphism group is an
infinite dimensional symmetry group of Einstein's vacuum
equations. The scale group is a symmetry group as well, but it
does not generalize to an infinite dimensional symmetry group
in the general vacuum case since the vacuum equations are not
conformally invariant. Hence the relative balance between the
scale and automorphism group is broken when one generalizes to
the case without symmetries. As in the finite dimensional
context one can regard sources as breaking the underlying
vacuum symmetries and presumably source contractions induce
symmetry hierarchies of the general Einstein field equations.
Furthermore, one can introduce symmetry hierarchies that split
the spatial diffeomorphism group into infinite dimensional and
finite symmetry groups that induce hierarchy structures.

Although the scale symmetry fails to generalize to an infinite
dimensional symmetry group, this does not mean that it is not
useful in the general inhomogeneous case. In particular, for
so-called asymptotically silent singularities the generic
asymptotic dynamics is expected to be asymptotically governed
by the silent
boundary~\cite{uggetal03,andetal05,limetal06,heietal09}; there,
the dynamics become local and spatial coordinates act as index
sets. This means that, pointwise, the scale group and the group
of spatial frame transformations play a similar role as
ScaleAut, but the actual symmetry transformation groups are
determined by the representation of the metric one chooses,
which in turn may depend on global topological issues. As
regards generic singularities it is worth pointing out that the
present analysis may be of relevance to other theories that
attempt to describe the Planck regime of the very early
universe or the interior of black holes, see
e.g.~\cite{dametal03,henetal07} and references therein.
%In this paper we have only discussed continuous automorphism
%transformations, but there also exist discrete ones. Discrete
%automorphisms induce discrete symmetries of Einstein's field
%equations; they are also connected with the
%existence of special models with fewer degrees of freedom,
%which sometimes, but not always, are associated with
%spacetimes with multiply transitive symmetry groups.
The generality of the mechanism we have described in this paper
suggests that there is room for further developments.%; we hope
%that this paper has provided some ideas and material that will
%be useful in future applications.

\subsection*{Acknowledgments}
We thank Bob Jantzen for a critical reading of an earlier draft.
We gratefully acknowledge the hospitality of the Mittag-Leffler Institute,
where this work was initiated.
CU is supported by the Swedish Research Council.

\bibliographystyle{plain}

\begin{thebibliography}{90}

%1
\bibitem{taub51} A.H. Taub.
\newblock Empty space-times admitting a three-parameter group
of  motions.
\newblock{\it Ann.\ Math.} {\bf 53} 472 (1951).%472-490

%2
\bibitem{uggetal03} C. Uggla, H. van Elst, J. Wainwright, and
    G.F.R. Ellis.
\newblock The past attractor in inhomogeneous cosmology.
\newblock {\it Phys.\ Rev.\ D} {\bf 68}~:~103502 (2003).

%3
\bibitem{rohugg05} N. R\"ohr and C. Uggla.
\newblock Conformal regularization of Einstein's field
    equations
\newblock {\it Class. Quant. Grav.} {\bf 22} 3775 (2005).%3775-3787

%4
\bibitem{andetal05} L. Andersson, H. van Elst, W.C. Lim, and C.
    Uggla.
\newblock  Asymptotic Silence of Generic Singularities.
\newblock {\it Phys.\ Rev.\ Lett.} {\bf 94} 051101 (2005).

%5
\bibitem{gar04} D.~Garfinkle.
\newblock  Numerical simulations of generic singularities.
\newblock Phys.\ Rev.\ Lett.\ {\bf 93} 161101 (2004).

%6
\bibitem{limetal06} W.~C.~Lim, C~.Uggla and J.~Wainwright.
\newblock  Asymptotic Silence-breaking Singularities.
\newblock Class.\ Quantum\ Grav.\ {\bf 23} 2607 (2006). %2607-2630

%7
\bibitem{heietal09} J.M. Heinzle, C. Uggla, and N. R\"ohr.
\newblock The cosmological billiard attractor.
\newblock {\it Adv.\ Theor.\ Math.\ Phys.} {\bf 13} 293 (2009).% 293-407

%8
\bibitem{waiell97} J. Wainwright and G.F.R. Ellis.
\newblock {\em Dynamical systems in cosmology}.
\newblock (Cambridge University Press, Cambridge, 1997).

%9
\bibitem{jan01} R.T. Jantzen.
\newblock Spatially Homogeneous Dynamics: A Unified
Picture.
\newblock in {\it Proc.\ Int.\ Sch.\ Phys.\ ``E. Fermi" Course LXXXVI on
``Gamov Cosmology"\/}, R. Ruffini, F. Melchiorri, Eds. (North
Holland, Amsterdam, 1987) and in  {\it Cosmology of the Early
Universe\/}, R. Ruffini, L.Z. Fang, Eds. (World Scientific,
Singapore, 1984).
\newblock arXiv:gr-qc/0102035.

%10
\bibitem{estetal68} F.B. Estabrook, H.D. Wahlquist, and C.G.
    Behr.
\newblock Dyadic analysis of spatially homogeneous world
models.
\newblock {\it J.\ Math.\ Phys.} {\bf 9} 497 (1968).%497-504

%11
\bibitem{rin00} H. Ringstr\"om.
\newblock Curvature blow up in Bianchi VIII and IX vacuum
spacetimes.
\newblock {\it Class.\ Quantum Grav.} {\bf 17} 713 (2000). %713-731

%12
\bibitem{rin01} H. Ringstr\"om.
\newblock The Bianchi IX attractor.
\newblock {\it Annales Henri Poincar\'e} {\bf 2} 405 (2001).% 405-500

%13
\bibitem{heiugg09a} J.M. Heinzle and C. Uggla.
\newblock A new proof of the Bianchi type IX attractor theorem.
\newblock {\it Class.\ Quant.\ Grav.} {\bf 26} 075015 (2009).

%14
\bibitem{heiugg09b} J.M. Heinzle and C. Uggla.
\newblock Mixmaster: Fact and Belief
\newblock {\it Class.\ Quant.\ Grav.} {\bf 26} 075016 (2009).

%16
\bibitem{jan79} R.T. Jantzen.
\newblock The dynamical degrees of freedom in spatially homogeneous
cosmology.
\newblock {\it Commun.\ Math.\ Phys.} {\bf 64} 211 (1979).%211-232

%16
\bibitem{rosetal90a} K. Rosquist, C. Uggla, and R.T. Jantzen.
\newblock Extended dynamics and symmetries in vacuum Bianchi
cosmologies.
\newblock {\it Class.\ Quant.\ Grav.} {\bf 7} 611 (1990).%611-624

%17
\bibitem{rosetal90b} K. Rosquist, C. Uggla, and R.T. Jantzen.
\newblock Extended dynamics and symmetries in perfect fluid Bianchi
cosmologies.
\newblock {\it Class.\ Quant.\ Grav.} {\bf 7} 625 (1990).%625-6637

%18
\bibitem{uggetal91} C. Uggla,  R.T. Jantzen, K. Rosquist, and
    H. von Z\"ur-Mühlen.
\newblock Remarks about late stage homogeneous cosmological
dynamics.
\newblock {\it Gen.\ Rel.\ Grav.} {\bf 23} 947 (1991).%947-966

%19
\bibitem{heietal05} J.M. Heinzle, N. R\"ohr, and C. Uggla.
\newblock Matter and dynamics in closed cosmologies.
\newblock {\it Phys.\ Rev.\ D} {\bf 71} 083506 (2005).

%20
\bibitem{mis69a} C.W. Misner.
\newblock Mixmaster universe.
\newblock {\it Phys.\ Rev.\ Lett.} {\bf 22} 1071 (1969).%1071-1074

%21
\bibitem{mis69b} C.W. Misner.
\newblock Quantum cosmology I.
\newblock {\it Phys.\ Rev.} {\bf 186} 1319 (1969).%1319-1327

%22
\bibitem{sik80} S.T.C. Siklos.
\newblock Field equations for spatially homogeneous spacetimes.
\newblock  {\it Phys.\ Lett.\ A} {\bf 76A} 19 (1980).%19-21

%23
\bibitem{sik81} S.T.C. Siklos.
\newblock Some Einstein spaces and their global properties.
\newblock  {\it J.\ Phys.\ A: Math.\ Gen.} {\bf 14} 395 (1981).%395-409

%24
\bibitem{chrter06} T. Christodoulakis and P.A. Terzis.
\newblock Automorphisms and a Cartography of the Solution Space
for Vacuum Bianchi Cosmologies: The Type III Case.
\newblock {\it J.\ Math.\ Phys.} {\bf 47} 102502 (2006).

%25
\bibitem{janugg99}  R.T. Jantzen and C. Uggla.
\newblock The kinematical role of automorphisms in the
orthonormal frame approach to Bianchi cosmology.
\newblock {\it J.\ Math.\ Phys.} {\bf 40} 353 (1999).%353-368

%26
\bibitem{stryor67} A. Strauss and J.A. Yorke.
\newblock On asymptotically autonomous differential equations.
\newblock {\it Math.\ Syst.\ Theory} {\bf 1} 175 (1967).%175-182

%27
\bibitem{ashetal93} A.~Ashtekar, R.~S.~Tate, and C.~Uggla.
\newblock Minisuperspaces: Observables and Quantization.
\newblock Int.\ J.\ Mod.\ Phys.\ {\bf D2} 15 (1993). %15-50

%28
\bibitem{ashetal93b} A.~Ashtekar, R.~S.~Tate, and C.~Uggla.
\newblock  Minisuperspaces: Symmetries and Quantization.
\newblock. In: Misner Festschrift, Edited by B.L. Hu et al.,
(Cambridge University Press (1993); arXiv:gr-qc/9302026.

%29
\bibitem{chrter07} T. Christodoulakis and P.A. Terzis.
\newblock The General Solution of Bianchi Type III Vacuum
Cosmology.
\newblock {\it Class.\ Quant.\ Grav.} {\bf 24} 875 (2007).%875-887

%30
\bibitem{terchr09} P.A. Terzis and T. Christodoulakis.
\newblock The general solution of Bianchi type VII$_h$ vacuum
cosmology.
\newblock {\it Gen. Rel. Grav.} {\bf 41} 469 (2009).%469-49

%31
\bibitem{rentod99} A.D. Rendall and K.P. Tod.
\newblock Dynamics of spatially homogeneous solutions of the
Einstein-Vlasov equations which are locally rotationally
symmetric.
\newblock {\it  Class.\ Quant.\ Grav.} {\bf 16} 1705
(1999).%1705-1726

%32
\bibitem{renugg00} A.D. Rendall and C. Uggla.
\newblock Dynamics of spatially homogeneous locally rotationally
symmetric solutions of the Einstein-Vlasov equations.
\newblock {\it Class.\ Quant.\ Grav.} {\bf 17} 4697 (2000).%4697-4714

%33
\bibitem{heiugg06} J.M. Heinzle and C. Uggla.
\newblock Dynamics of the spatially homogeneous Bianchi type I
Einstein-Vlasov equations.
\newblock {\it Class.\ Quant.\ Grav.} {\bf 23} 3463 (2006).%3463-3490

%34
\bibitem{dametal03} T. Damour, M. Henneaux,  and H. Nicolai.
\newblock Cosmological billiards.
\newblock {\it Class.\ Quant.\ Grav.} {\bf 20} R145 (2003).%R145-R200

%35
\bibitem{henetal07} M. Henneaux, D. Persson, P. Spindel.
\newblock Spacelike Singularities and Hidden Symmetries of
Gravity.
\newblock {\it Living Rev.\ Rel.} {\bf 11} 1 (2008);
arXiv:0710.1818.


%%%%%%%%%%%%%%%%%%%%%%%%%%%%%%%%%%%%%%%%%%%%%%%%%%%%%%%%%%%%%
%%%%%%%%%%%%%%%%%%%%%%%%%%%%%%%%%%%%%%%%%%%%%%%%%%%%%%%%%%%%%
%%%%%%%%%%%%%%%%%%%%%%%%%%%%%%%%%%%%%%%%%%%%%%%%%%%%%%%%%%%%%

\end{thebibliography}

\vfill
%\pagebreak

\end{document}